\documentstyle[prb,aps,epsfig,amsfonts]{revtex}

\begin{document}

\twocolumn \psfull \draft

\wideabs{
\title{Equilibrium Properties of Double-Screened-Dipole-Barrier
SINIS Josephson Junctions}
\author{Branislav K. Nikoli\' c$,^1$ J. K. Freericks,$^1$ and P. Miller$^2$}
\address{$^1$Department of Physics, Georgetown University,
Washington, DC 20057-0995 \\
$^2$Department of Physics, Brandeis University, Waltham, MA 02454}

\maketitle

\begin{abstract}
We report on a self-consistent microscopic study of the DC
Josephson effect in $SINIS$ junctions where screened dipole
layers at the $SN$ interfaces generate a double-barrier
multilayered $SIN$ structure. Our approach starts from a
microscopic Hamiltonian defined on a simple cubic lattice, with
an attractive Hubbard term accounting for the short coherence
length superconducting order in the semi-infinite leads, and a
spatially extended charge distribution (screened dipole layer)
induced by the difference in Fermi energies of the superconductor
$S$ and the clean normal metal interlayer $N$. By employing the
temperature Green function technique, in a continued fraction
representation, the influence of such spatially inhomogeneous
barriers on the proximity effect, current-phase relation, critical
supercurrent and normal state junction resistance, is
investigated for different normal interlayer thicknesses and
barrier heights. These results are of relevance for high-$T_c$
grain boundary junctions, and also reveal one of the mechanisms
that can lead to low critical currents of apparently ballistic
$SNS$ junctions while increasing its normal state resistance in a
much weaker fashion. When the $N$ region is a doped semiconductor,
we find a substantial change in the dipole layer
(generated by a small Fermi level mismatch) upon crossing 
the superconducting critical temperature, which is a new signature 
of proximity effect and might be related to recent Raman 
studies in Nb/InAs bilayers.
\end{abstract}

}

\narrowtext

\section{Introduction}

The Josephson effect~\cite{josephson} is one of the most
spectacular phenomena arising from the macroscopic phase
coherence of Cooper pairs. A dissipationless current flows at zero
voltage between two superconductors weakly coupled through a
tunnel barrier ($SIS$, where $S$ and $I$ denote a superconductor and
an insulating barrier, respectively) or weak links ($ScS$, $SNS$,
etc., where $c$ stands for a constriction, and $N$ for a normal
metal). The study of such inhomogeneous superconducting
structures has been driven by both interest in the fundamentals of
quantum mechanics, and by the potential application of Josephson
junctions as circuit elements in electronic
devices.~\cite{vanduzer}

Recently, considerable attention has been directed toward the
study of $SINIS$ junctions,~\cite{kleinssaser,brinkman,volkov}
where the insulating tunnel barrier is split into two pieces
separated by a normal metal. These types of junctions have
provided a playground to study the interplay~\cite{carlo_rmt}
between the mesoscopic coherence of a single-particle wave
function in the normal metal and the macroscopic coherence of a
many-body wave function of Cooper pairs.~\cite{superlattices}
Furthermore, the reexamination of various multilayered structures
of the $SINIS$ type in applied research  has been driven by the
struggle to optimize the performance of Josephson junctions in
low-temperature superconducting (LTS) digital
electronics.\cite{maezawa,sinis_review,rusi_sct} In mesoscopic
superconductivity, one frequently deals with $S$-$Sm$-$S$
junctions~\cite{kleinssaser} ($Sm$ being a heavily doped
semiconductor with a two-dimensional electron gas) where the role
of the $I$ layer is played by a space-charge layer arising at the
$S$-$Sm$ interface (additional scattering at the interface can
occur from the mismatch between the effective electron masses and
Fermi momenta in the $S$ and $Sm$). The technological advances in
fabricating such hybrid structures~\cite{kleinssaser} have given
an impetus to the field of mesoscopic
superconductivity~\cite{carlo_rmt,superlattices} where the
two-dimensional electron gas is amenable to an engineering of its
``metallic'' properties; i.e, one can tune the Fermi wavelength,
or mean free path, and one can confine electrons with gate
electrodes. In such structures,  phase-coherence of the electron
and Andreev-reflected hole~\cite{andreev} at the $SN$ interface
can be studied without too much normal reflection, because the
charge-accumulation layer arising at a typical Nb/InAs interface,
or the Schottky barrier at a
Nb/Si interface, are much more transparent than typical dielectric tunnel
barriers.~\cite{carlo_rmt}

While initial understanding of the Josephson effect came from
studies of tunnel junctions,~\cite{josephson} further
developments concentrated on weak links~\cite{likharev_review}
which provide the non-hysteretic (i.e., single valued) $I-V$
characteristic needed for applications, like SQUIDs~\cite{squid}
or rapid single flux quantum logic.~\cite{rsfq} The return to
$SIS$ junctions came after the fabrication of Nb/Al tunnel
junctions~\cite{gurvitch} with a reliable control of the critical
current (conventional tunnel junctions can be made non-hysteretic
by externally shunting their high capacitance with a resistor,
which reduces the overall performance~\cite{klein_ieee}). The
renewed interest~\cite{sinis_review} in $SINIS$ multilayered
junctions for LTS electronics comes from an attempt to combine
the advantageous properties of  both weak links and tunnel
junctions~\cite{maezawa}---the $SINIS$ junctions are
intrinsically shunted, while exhibiting large characteristic
voltages with moderate critical current densities (in fact, rapid
single flux quantum devices require large critical current
densities, to reduce the error rate,~\cite{rsfq} which is difficult
to achieve using standard Nb/Al/AlO$_x$Nb tunnel junction
technology, but might be reached in $SINIS$ junctions with
carefully engineered properties~\cite{sinis_review}). When the
$N$ interlayer is clean, the junction resistance is mainly controlled by
scattering at the interfaces (like in conventional
Nb/Al/${\rm AlO_x}$/Al/Nb junctions~\cite{zehnder}), and not by
the interlayer material properties.

Here we undertake a study of a special class of $SINIS$ junctions
where the double-barrier structure arises from two inhomogeneous
screened dipole layers (SDL) determined by a relatively large
Debye screening length $l_D$ of a few lattice spacings. We start
from a microscopic lattice Hamiltonian with the $S$ and $N$ layers
described by different metals that have the same bandwidth, but
their Fermi levels are misaligned. The Fermi level mismatch
forces a charge redistribution, with the strongest deviation from
uniformity located near the $SN$ interface, which is gradually
diminished inside the bulk layers on a length scale set by
$l_D$. The charge profile ensures an equilibration of the
chemical potential throughout the system when no bias voltage is
applied. Since we assume a screening length of a few lattice
spacings, the dipole layer is spatially extended (i.e., thicker
than just one monoatomic layer). This choice of microscopic
junction parameters allows us to examine the charge
redistribution appearing between conductors which are less
efficient in screening than ordinary metals (such as the
underdoped cuprates or InAs). Our treatment of the double SDL
barrier is fully microscopic and self-consistent, meaning that
effects of the static electric potential (generated by the
excess charge) on the Josephson current and on  the normal state
resistance are related to the parameters of the underlying
Hamiltonian, rather than characterizing the barrier by an
effective transparency~\cite{brinkman,rusi_sct,lukichev} $D$, or
using a delta function potential at the $SN$ interface to model
the normal reflection~\cite{blonder,furusaki} (in addition to the
inevitable retroreflection~\cite{andreev}). We tackle both the
fundamental aspects of the problem (like the self-consistent
evaluation of the order parameter, the change of its phase across
the junction, and the emergence of non-sinusoidal current-phase
relations) and issues relevant for applications (like the
characteristic voltage, a product of the critical current $I_c$
and the normal state resistance $R_N$, which determines the
high-frequency performance of the junction). Our junctions are
three-dimensional (3D) and clean, so that quasiparticle transport
through the $N$ interlayer is ballistic.

Previous theoretical work on ballistic $SINIS$ junctions focused
on resonant supercurrents  in low-dimensional
structures.~\cite{furusaki,chrestin,wendin,ivan} Mesoscopic
superconductivity coherence effects  in 3D junctions (e.g., a
current proportional to  $D$ of the barrier, rather than the
characteristic $D^2$ dependence for two uncorrelated sequential
tunneling processes) have been investigated in
Ref.~\onlinecite{brinkman}. These junctions are mostly similar to
the ones studied here, except that our ``microscopic'' charge
accumulation barriers are not atomically sharp interfaces that
can be described by a phenomenological transparency $D$. A more
microscopic treatment of the effect of charge inhomogeneity for 
\begin{figure}
\centerline{\psfig{file=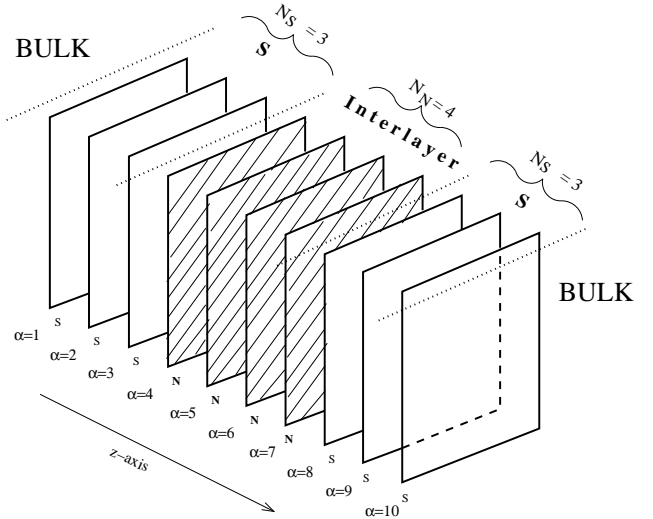,height=3.0in,angle=0} }
\vspace{0.2in} \caption{ Microscopic stacked planar geometry of a
Josephson junction defined on an infinite simple cubic lattice
with a lattice constant $a$. The normal interlayer contains $N_N$
planes (ranging from 1 to 60)  which are coupled to semi-infinite
superconducting leads (the junction thickness is $L=N_Na$). These
layers, together with the first $N_{S}$ planes (30 in our
calculations) in each lead, comprise the region of the junction
where the self-consistent calculation is performed. The junction
is allowed to have spatial inhomogeneity only within the
$2N_{S}+N_N$ modeled planes, but the calculations are for an
infinite system. The insulating barriers are formed by a charge
redistribution that is localized near the $SN$ interfaces.} \label{fig:planes}
\end{figure}
normal transport through the contact of two different metals (a
problem frequently appearing in the multilayers of giant
magnetoresistance devices~\cite{gijs}) has been undertaken using
the Boltzmann equation,~\cite{dugaev} and in superconducting
junctions using quasiclassical methods in a non-self-consistent
fashion.~\cite{zaitsev} It is worth emphasizing that standard
quasiclassical Green function techniques, which exploit the fact
that macroscopic quantities vary
on a length scale substantially exceeding the interatomic distance, cannot be
applied directly to problems containing boundaries between
two different metals. Since electron reflections lead to fast
spatial variations of the original Green functions around the
boundary, the method has to be extended properly to take this
into account (see Ref.~\onlinecite{zaitsev} for details).

Our study is relevant for three types of recently explored
experimental systems: (i) grain boundary junctions~\cite{delin} in
high-$T_c$ superconductors, where our short coherence length
superconductor and the poor screening of the excess charge (i.e.,
Debye screening length comparable to the coherence length), mimic
the effect of a charge imbalance at the grain boundaries on the
depression of the order parameter, and thereby the intergranular
current density~\cite{mannhart98,gurevich98} (without
complicating the problem further with $d$-wave symmetry); (ii)
Raman scattering studies~\cite{igor} of the proximity effects in
Nb/InAs hybrid structures reveal a substantial change of the
charge accumulation layer formed at such interface above and
below the $T_c$ of Nb---we also find that $I$ layer induced by a small
Fermi level mismatch is modified by proximity effects in our
$SINIS$ junctions when the carrier concentration in the $N$ is 100
times smaller than in the $S$; (iii) recent experiments on
ballistic $SNS$ junctions,~\cite{klapwijk} in the limit where
$I_c$ and $R_N$ do not depend on the thickness of the $N$,
exhibit a much smaller characteristic voltage than predicted for
short clean $ScS$ junctions---the scattering off a dipole charge
layer is an example of a process which sharply reduces $I_c$, but
only weakly increases $R_N$.

The paper is organized as follows. In Sec.~\ref{sec:lattice} we
introduce the model and the main ideas of the Green function
computational technique (employed to solve the quantum problem of
the charge distribution and equilibrium transport; the electrostatic
problem of the potential generated by
these charges is solved classically).
Section~\ref{sec:phase} contains the results for
the self-consistent pair amplitude (or the order parameter) and the
local change of the phase across the junction. The current-phase
relation for different strengths of the electrostatic potential
generated by the SDL is discussed in Sec.~\ref{sec:icrn}, where
we also evaluate the characteristic voltage $I_cR_N$. We
conclude in Sec.~\ref{sec:conclusion}.

\section{Modeling a $SINIS$ Junction with
a  double-barrier screened dipole layer}\label{sec:lattice}

Early studies of the Josephson effect in $SINIS$ junctions were
based on a tunneling Hamiltonian formalism and perturbation
theory in the barrier transmissivity.~\cite{larkin} Later on,
quasiclassical Green function techniques~\cite{schon} were
applied to a double-barrier junction with the $N$ interlayer in the
dirty limit.~\cite{lukichev} While these results are valid only
in a few limiting cases, a recent reexamination of this problem
covers a wider range of parameters.~\cite{brinkman,sinis_review}
For example, when transport through the $N$ interlayer is
ballistic (mean free path greater than the thickness of the
junction), one cannot use standard tools~\cite{lukichev} like
the Usadel equation. Instead, a solution of the Gor'kov equations
for the Green functions of the double-barrier structure is
required.~\cite{brinkman,rusi_sct} Furthermore, if the $I$
barriers are not of low transparency, the usual arguments for the
validity of rigid boundary conditions~\cite{likharev_review}
(i.e., taking the gap $\Delta$ to be constant inside the superconducting leads)
fail when the $S$ and $N$ regions have the same cross section, and the
thickness of the junction is not much larger than the
superconducting coherence length $\xi_0$. In such cases, the
critical current density can be close to the bulk critical
current density, and a self-consistent evaluation of the order parameter
inside the $S$ is needed to ensure current conservation throughout
the structure.~\cite{levy,sols,bagwell} Since we choose to work
with a short coherence length superconductor, quasiclassical
approximations neglecting dynamics on a length scale below
$\xi_0$ are not applicable (in our case $\xi_0 \approx 4a$ is not much
larger than the Fermi wavelength $\lambda_F \approx 2a$, and spatial variation
of the order parameter $\Delta$ on a length scale smaller or
comparable to $\xi_0$ is important).

Our approach to quantum transport in ballistic $SINIS$ junctions
starts from a microscopic Hamiltonian defined on a simple cubic
lattice (of lattice constant $a$).~\cite{levy} It allows us to describe
the transport for
an arbitrary junction thickness, temperature, and barrier strength.
Also, the geometry is such that the $N$ interlayer has the same width
as the $S$ leads. For computational purposes, the infinite lattice
which models the junction is divided into a self-consistent
part and a bulk part, as shown in Fig.~\ref{fig:planes}.
A negative-$U$ Hubbard term is employed to model the
real-space pairing of electrons due to a local instantaneous attractive
interaction.~\cite{levy,annett} The lattice Hamiltonian is given
by
\begin{eqnarray} \label{eq:tbh}
H & = & \sum_{i\sigma} V_i c_{i\sigma}^{\dag}c_{i\sigma} - \sum_{\langle
ij\sigma
\rangle }t_{ij}c_{i\sigma}^{\dag}c_{j\sigma} \nonumber \\
&&  + \sum_i U_i\left (
c_{i\uparrow}^{\dag}c_{i\uparrow}-\frac{1}{2}\right ) \left (
c_{i\downarrow}^{\dag}c_{i\downarrow}-\frac{1}{2}\right ),
\end{eqnarray}
where $c_{i\sigma}^{\dag}$ ($c_{i\sigma}$) creates (destroys) an
electron of spin $\sigma$ at site $i$, $t_{ij}$ is the
hopping integral between nearest-neighbor sites
$i$ and $j$ (energies are measured in units of $t$), which is taken
to be the same in the $S$ and $N$, and $U_i < 0$ is the attractive
Hubbard interaction for sites
within the superconducting planes. The normal interlayer is described
by the noninteracting part of the Hamiltonian~(\ref{eq:tbh}), which is
just a (clean) nearest-neighbor tight-binding model with
a diagonal on-site potential $V_i$.
The potentials $V_i$ are not given {\em a priori}, but instead are
calculated self-consistently by first determining the local
electronic charge density and  comparing it to the bulk charge
density of the corresponding $S$ or $N$ layers. The imbalanced
charge on each plane generates an electric field and thereby an
electric potential. Summing the contributions from the  charges on all
other planes then yields the total local potential $V_i^C$ and the
local potential energy shift $V_i=eV_i^C$. We now recalculate
the charge density on each plane and iterate until $V_i$ is
determined self-consistently (see below for a detailed
description of the algorithm). The local potentials $V_i$ are
largest near the $SN$ interface, and decay as
one approaches the bulk leads.

We use the Hartree-Fock approximation (HFA) for the interacting
part of the Hamiltonian~(\ref{eq:tbh}). This accounts for the
superconductivity in the $S$ region in a way which is completely
equivalent to a conventional BCS theory with an energy cutoff
determined by the electronic bandwidth rather than by the phonon
frequency. We choose half-filling $n_S=1$ and $U_i=-2$
on the sites in the superconducting leads. The homogeneous bulk
superconductor has a transition temperature $T_c=0.11$ and a zero-temperature order
parameter $\Delta=0.198$. This yields a standard BCS gap ratio
$2\Delta/(k_BT_c)\approx 3.6$ and a short 
\begin{figure}
\centerline{\psfig{file=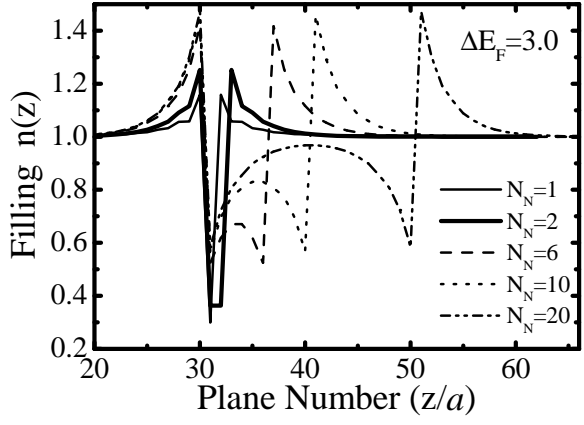,height=3.0in,angle=-90} }
\vspace{0.2in} \caption{Scaling of the local density of electrons
 with the thickness $L=N_Na$ of $SINIS$ Josephson
junctions. The filling is the same at each site within the planes
(plane number one is the first plane along the $z$-axis inside
the self-consistent region from Fig.~\ref{fig:planes}). The
difference in Fermi energies of the $S$ and $N$ is $\Delta
E_F=3.0$. The bulk equilibrium value of the charge density is set
to half-filling in both the $S$ and $N$ ($n_S=n_N=1$). The
inhomogeneous charge redistribution is (approximately) symmetric
around the SN interface only for a thick enough junction. In the case
of a thin junction, the charge is depleted in the $N$ interlayer to lie
below half-filling since the screening length is a few lattice
spacings.} \label{fig:n_def3}
\end{figure}
coherence length
$\xi_0=\hbar v_F/(\pi\Delta) \simeq 4a$. The bulk critical
current per unit area $a^2$ is $I_c^{\rm bulk}=1.09 e n
\Delta/\hbar k_F$, which is a bit higher than the current density
determined by the Landau depairing velocity $v_d=\Delta/\hbar
k_F$. This stems from  the possibility of having gapless
superconductivity in 3D at superfluid velocities slightly
exceeding~\cite{bardeen} $v_d$ (note that $k_F$ is
direction-dependent for a cubic lattice at half-filling; we use
the average value over the Fermi surface $k_F \approx 2.8a$,
appearing in the transport formulas, to compare our critical bulk
supercurrent density to the expressions that assume a spherical
Fermi surface and a density of particles $n=k_F^3/3\pi^2$). The
junction properties are studied here in the low-temperature limit
at $T=0.01=0.09T_c$ (the BCS gap is essentially temperature
independent below $0.6T_c$). At this temperature, the
coherence length of the clean normal metal is
$\xi_N=\hbar v_F/2\pi kT \simeq 40a$. Since we do not
consider inelastic scattering processes, the dephasing length
$L_\phi$ is larger than $\xi_N$. Therefore, ${\rm min} \,(\xi_N,L_\phi)=\xi_N$
determines the coherence properties of a single quasiparticle wave
function inside the normal region.

The inhomogeneous superconductivity problem is solved by
employing a Nambu-Gor'kov matrix formulation for the Green
function $\hat{G}({\bf r}_i,{\bf r}_j,i\omega_n)$ between two
lattice sites ${\bf r}_i$ and ${\bf r}_j$ at the Matsubara
frequency $i\omega_n=i\pi T(2n+1)$,
\begin{equation}
\hat{G}({\bf r}_{i},{\bf r}_{j},i\omega_{n}) = \left( \begin{array}{cc}
G({\bf r}_{i},{\bf r}_{j},i\omega_{n}) & F({\bf r}_{i},{\bf
r}_{j},i\omega_{n}) \\
\overline{F}({\bf r}_{i},{\bf r}_{j},i\omega_{n}) &
- G^{*}({\bf r}_{i},{\bf r}_{j},i\omega_{n})
\end{array} \right).
\label{eq:g_nambu}
\end{equation}
\begin{figure}
\centerline{\psfig{file=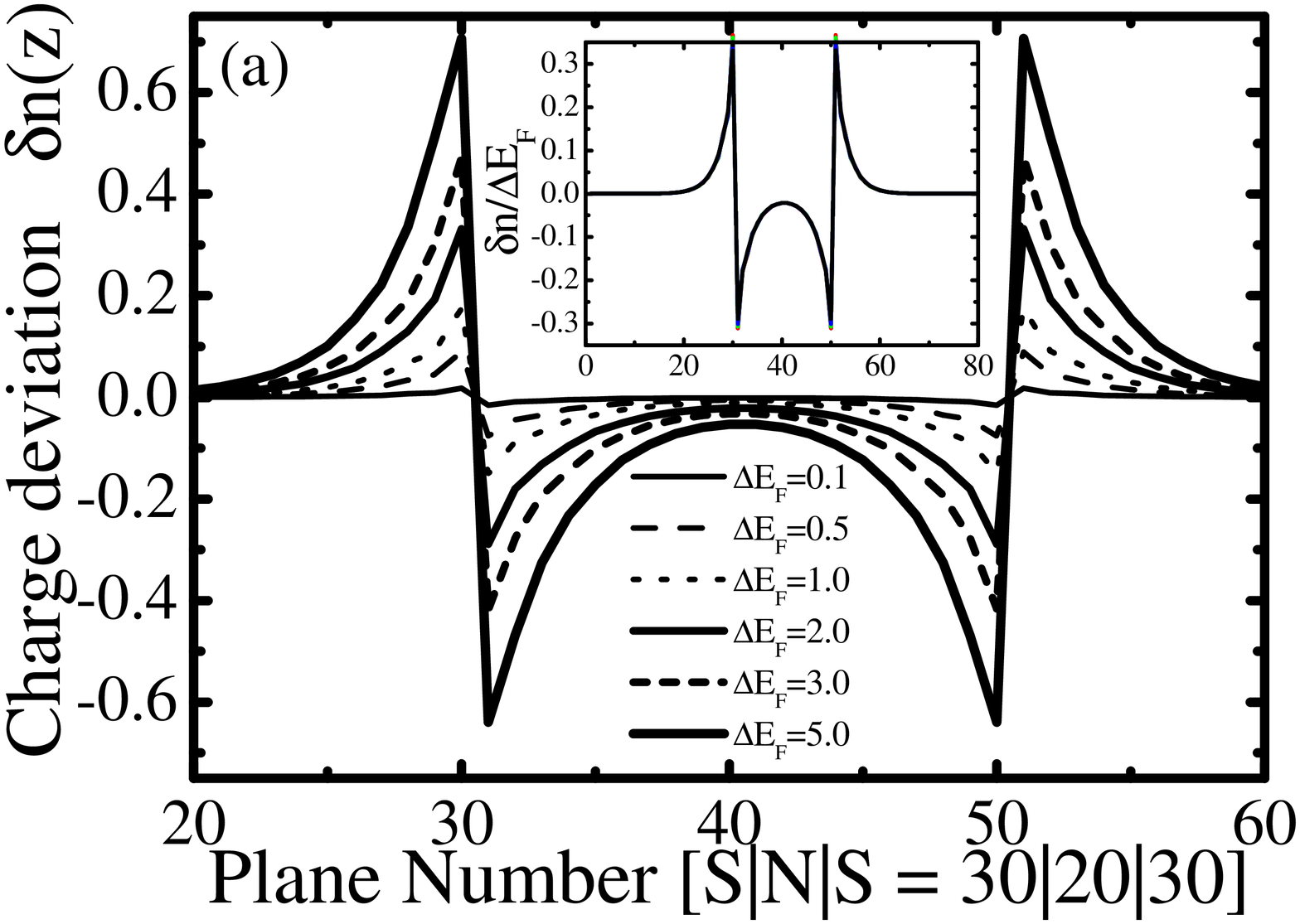,height=3.0in,angle=-90} }
\vspace{0.2in}
\centerline{\psfig{file=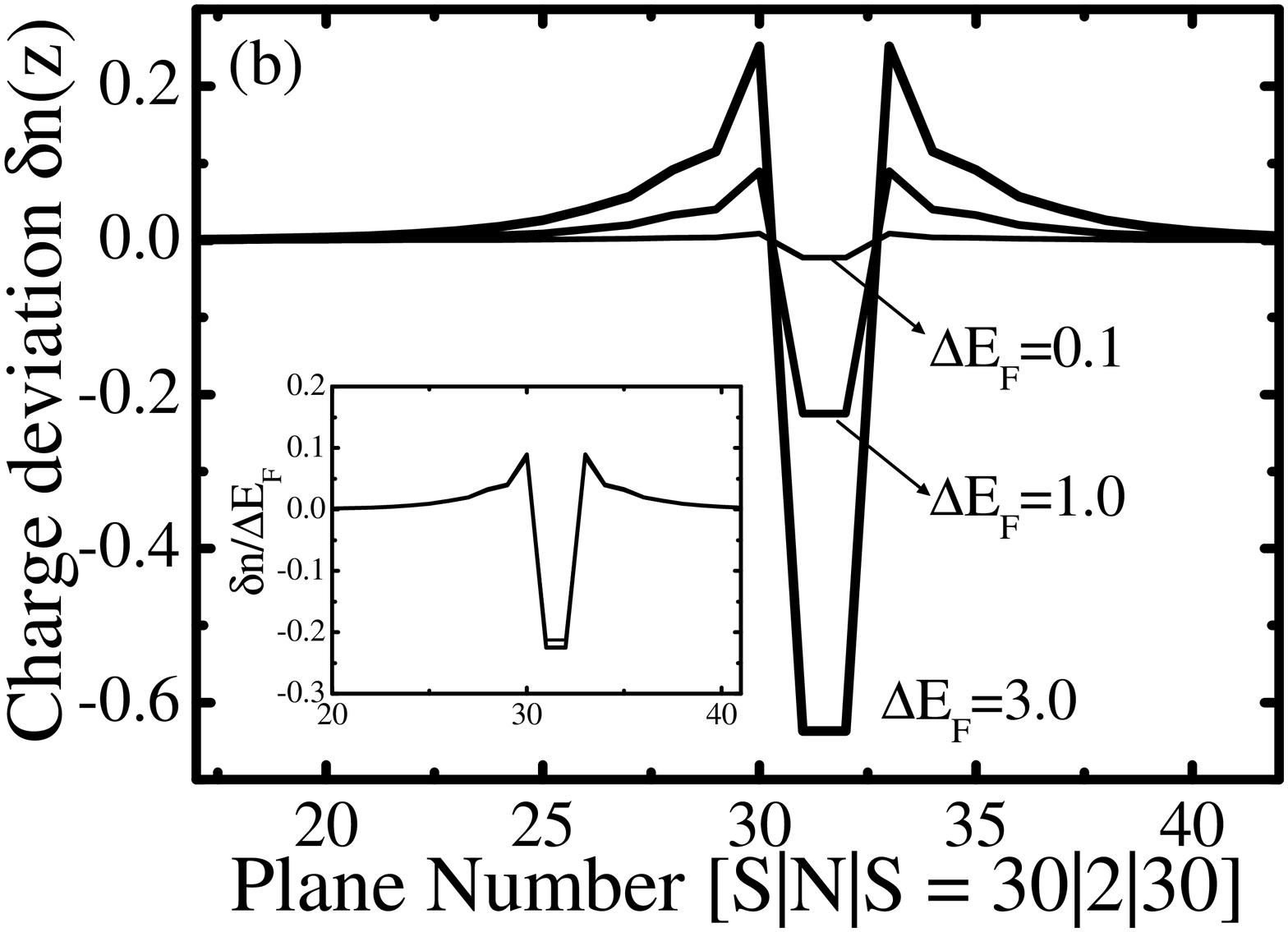,height=3.0in,angle=-90} }
\vspace{0.2in} \caption{Charge deviation $\delta n(z)$ (from
half-filling) in $SINIS$ junctions characterized by different
Fermi level mismatches $\Delta E_F=E_F^N-E_F^S$. The $N$ interlayer
consists of: (a) 20 normal planes (b) 2 normal planes. The inset
shows that distributions of $\delta n(z)$ for different $\Delta
E_F$ can be rescaled to a single curve after multiplying them by
the ratio $(\Delta E_F)_{\rm ref}/\Delta E_F$, where $(\Delta
E_F)_{\rm ref}=1.0$ is chosen as the reference distribution.} \label{fig:dn_L20}
\end{figure}
The corresponding local self-energy is given by the matrix
\begin{equation}
\hat{\Sigma} ({\bf r}_{i},i\omega_{n}) = \left( \begin{array}{cc}
\Sigma({\bf r}_{i},i\omega_{n}) & \phi({\bf r}_{i},i\omega_{n})
\\ \phi^{*}({\bf r}_{i},i\omega_{n}) & - \Sigma^{*}({\bf
r}_{i},i\omega_{n})
\end{array} \right).
\label{eq: sigma_nambu}
\end{equation}
The diagonal and off-diagonal (i.e., normal and anomalous) Green
functions are defined, respectively, as
\begin{eqnarray}
G({\bf r}_{i},{\bf r}_{j},i\omega_{n}) & = & -  \int\limits^{\beta}_{0}
d\tau \exp (i\omega_{n}\tau) \langle{\rm T}_{\tau}
\hat{c}_{j\sigma}(\tau) \hat{c}^{\dagger}_{i\sigma}(0) \rangle,
\\
F({\bf r}_{i},{\bf r}_{j},i\omega_{n}) & = & - \int\limits^{\beta}_{0}
d\tau \exp (i\omega_{n}\tau) \left\langle{\rm T}_{\tau}
\hat{c}_{j\uparrow}(\tau) \hat{c}_{i\downarrow}(0) \right\rangle,
\label{eq: g_f_def}
\end{eqnarray}
where ${\rm T}_{\tau}$ denotes time-ordering in $\tau$ and
$\beta=1/T$. The self-energies and Green functions are coupled
together through the Dyson equation,
\begin{figure}
\centerline{\psfig{file=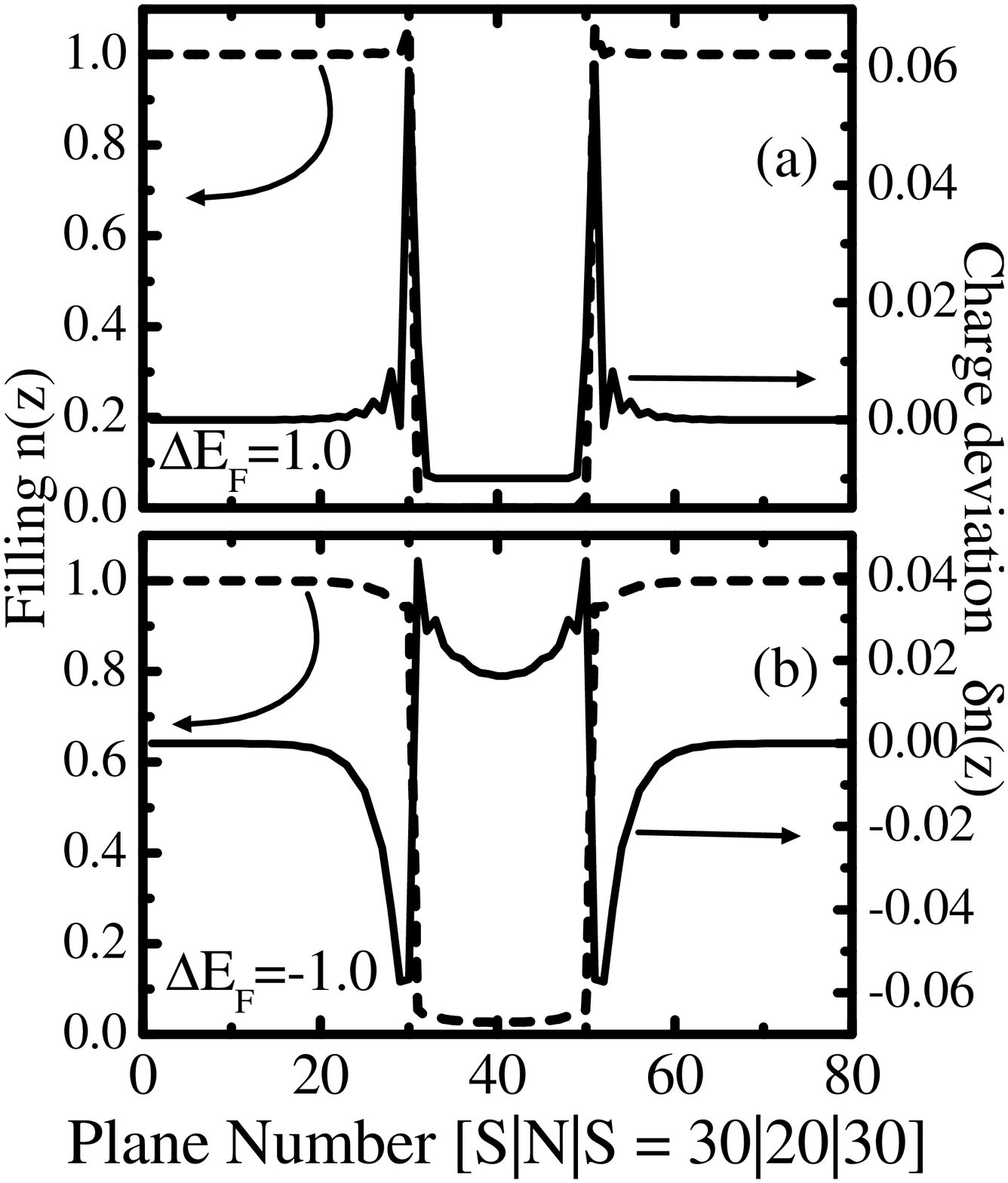,height=3.0in,angle=0} }
\vspace{0.3in}
\centerline{\psfig{file=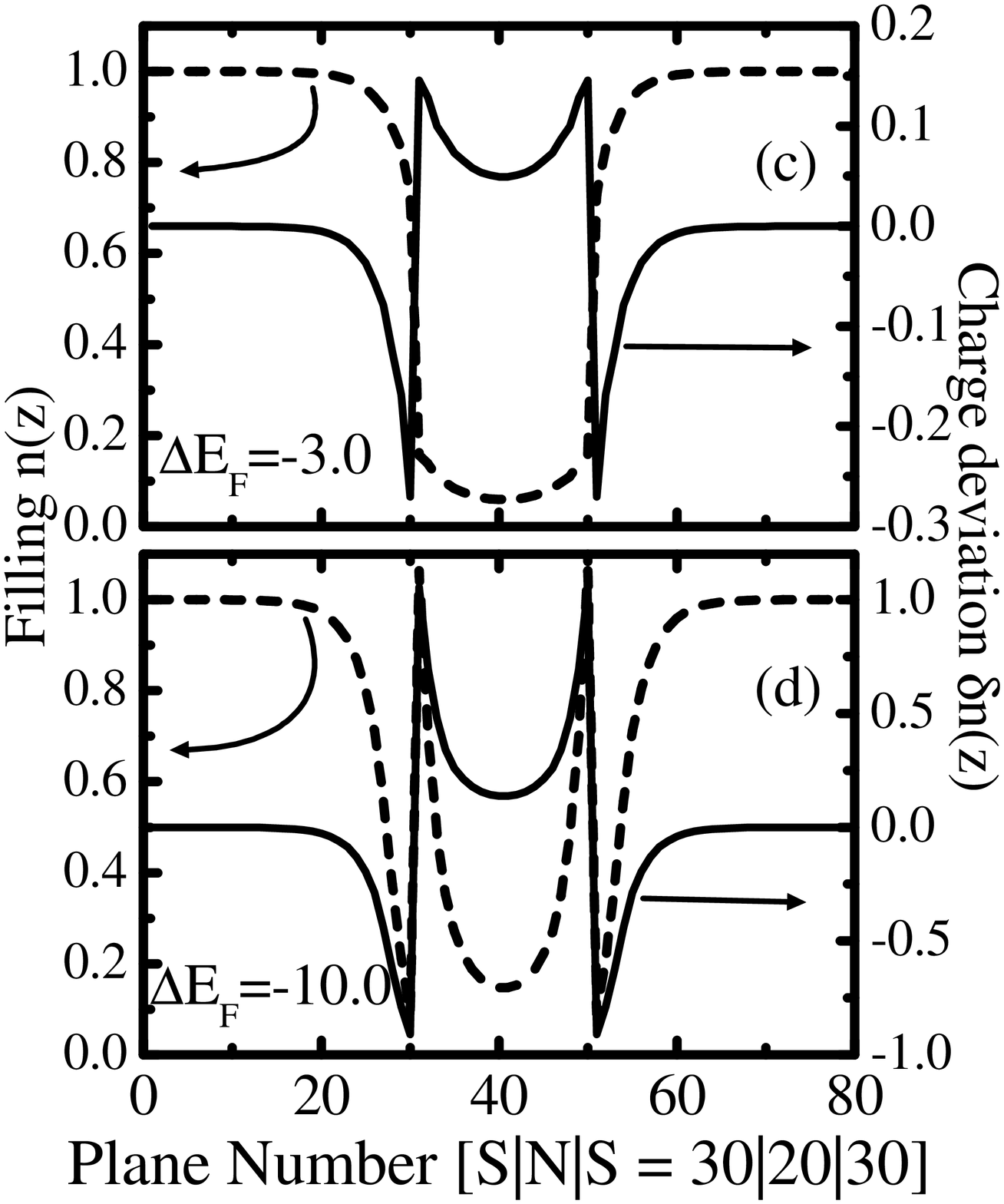,height=3.0in,angle=0} }
\vspace{0.2in} \caption{Electron filling $n(z)$ [dashed line] and
charge deviation $\delta n(z)$ [solid line] of a $SINIS$
Josephson junction with a  thickness $L=20a$ and the $N$
interlayer chosen to approximate a highly doped semiconductor. The
charge deviation is measured with respect to the equilibrium
filling in the bulk, $n_S=1$ in the superconductor and $n_N=0.01$
in the normal region. The Fermi energy mismatch $\Delta
E_F=E_F^N-E_F^S$ between the $N$ and the $S$ is: (a) $\Delta E_F =
1.0$, (b) $\Delta E_F = -1.0$, (c) $\Delta E_F = - 3.0$, and (d)
$\Delta E_F = - 10.0$. The charge profile is virtually
independent of temperature, both above and below $T_c$.}
\label{fig:n001_def1}
\end{figure}
\begin{eqnarray}
\lefteqn{\hat{G}({\bf r}_{i},{\bf r}_{j},i\omega_{n})= G^0({\bf
r}_{i},{\bf r}_{j},i\omega_{n})} \nonumber \\
& & \mbox{} +
\sum_{l} G^0({\bf r}_{i},{\bf r}_{l},i\omega_{n}) \hat{\Sigma}({\bf
r}_{l},i\omega_{n}) \hat{G}({\bf r}_{l},{\bf r}_{j},i\omega_{n}),
\label{eq: Dys}
\end{eqnarray}
where the local approximation for the self-energy, $\Sigma({\bf
r}_{i},{\bf r}_{j},i\omega_{n}) = \Sigma({\bf
r}_{i},i\omega_{n})\delta_{ij}$, is assumed. In the HFA
\begin{figure}
\centerline{\psfig{file=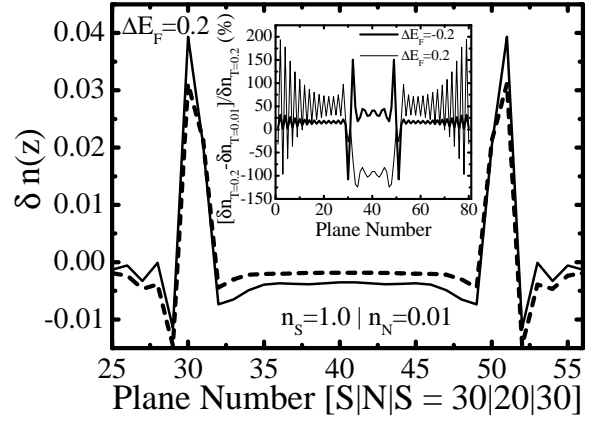,height=3.0in,angle=-90} }
\vspace{0.2in} \caption{Evolution of charge accumulation region 
in a $SINIS$ junction, with a highly doped semiconductor as 
the $N$ interlayer ($n_N=0.01$, $n_S=1.0$, $\Delta E_F=0.2$), upon crossing 
$T_c$ of the $S$ by going from $T=0.2=1.8T_c$ to $T=0.01=0.09T_c$. The inset 
shows the relative change of the charge deviation $[\delta n(z)_{T=0.2}-
\delta n(z)_{T=0.01}]/\delta n(z)_{T=0.2}$ for two different charge 
redistributions: $\Delta E_F=0.2$ and $\Delta E_F=-0.2$.}
\label{fig:change_dn}
\end{figure}
for the attractive Hubbard model, the local self-energy is found
from the local Green function by
\begin{equation}
\Sigma({\bf r}_{i},i\omega_{n}) = U_i T \sum_{\omega_{n}} G({\bf
r}_{i},{\bf r}_{i},i\omega_{n}),
\label{eq:selfa}
\end{equation}
and
\begin{equation}
\phi({\bf r}_{i},i\omega_{n}) = -U_i T \sum_{\omega_{n}} F({\bf
r}_{i},{\bf r}_{i},i\omega_{n}). \label{eq:selfb}
\end{equation}
The self-energy is time-independent because the interaction is
instantaneous and we use the HFA (i.e., retardation effects in
the superconductor are neglected).
The noninteracting Green
function, $G^0({\bf r}_{i},{\bf r}_{j},i\omega_{n})$ is diagonal
in Nambu space, with an upper diagonal component given by
\begin{equation}
G^{0}({\bf r}_{i},{\bf r}_{j},i\omega_{n}) = \int d^{3}{\bf k}
\frac{ \mbox{\rm e}^{i {\bf k}\cdot({\bf r}_{i}-{\bf r}_{j}) } }
{ i\omega_{n} +\mu-\varepsilon_{{\bf k}} }.
\end{equation}
As all sites within a plane are identical, the self-energy need
only be calculated once for each of the planes, while it is
allowed to vary from plane to plane.

We work with Green functions $\hat{G}_{\alpha, \beta}
(i\omega_{n},k_{x},k_{y})$ represented in a mixed basis, which is
defined by the two-dimensional momenta $(k_x,k_y)$ and the
(discrete) $z$-coordinate of the plane $\alpha=z_i/a$. This
follows after the initial 3D problem is converted to a
quasi-one-dimensional one~\cite{pott} by performing a Fourier
transformation within each plane (where the junction is
translationally invariant) and retaining the real-space
representation for the $z$-direction of the inhomogeneity. For the
local interaction treated in the HFA, computation of the Green
function reduces to inverting an infinite block tridiagonal
Hamiltonian matrix in real space. The Green functions are thereby
evaluated as a matrix continued fraction (technical details are
given elsewhere~\cite{miller,freericks}). The final solution is
fully self-consistent in the order parameter $|\Delta(z)|e^{i
\phi(z)}$ inside the part of the junction comprised of the $N$
region and the first 30 planes inside the superconducting leads
on each side of the $N$ interlayer (see Fig.~\ref{fig:planes}).
The self-consistent region is long enough because $|\Delta(z)|$
heals to its bulk value over just a few coherence lengths $\xi_0$.
Our Hamiltonian formulation of the problem and its solution by
this Green function technique is equivalent to solving a
discretized version of the Bogoliubov-de Gennes~\cite{dege} (BdG)
equations in a fully self-consistent manner, i.e. by determining
the off-diagonal pairing potential $\Delta_i$ in the BdG
Hamiltonian~\cite{levy} after each iteration until convergence is
achieved. The tight-binding description of the electronic states
also allows us to include an arbitrary band structure or
unconventional pairing symmetry.~\cite{annett}

In conjunction with the self-consistent solution of the
superconducting part of the problem, we have to self-consistently
solve the electrostatic problem. Although both the $S$ and $N$
are half-filled in most of our calculations (i.e., there is no
mismatch in the Fermi wave vector), shifting the bottom of the
$N$ band leads to a difference in their Fermi levels. This
generates a redistribution of electrons around the $SN$ interface
when these are brought into contact. The resulting non-uniform
electric field can be described by a potential $V(z)$ (for
simplicity, we use the label $z$ having in mind a discrete $z_i$
coordinate at a particular site $i$) which varies in the
transition layer around the $SN$ boundary with a thickness of
$2d$. In the region $z > |d|$ the following condition is satisfied
\begin{equation}
E_F^S+V(z<-d)=E_F^N+V(z> d)=\mu,
\end{equation}
in order to ensure a constant electrochemical potential $\mu$
throughout the system in equilibrium. The solution which
satisfies this equation is usually simplified~\cite{zaitsev} to
$V(z)=V_0\delta(z) + \mu - v(z)$, where $v(z)$ is a monotonic
function of $z$ equal to $E_F^S$ for $z<-d$ or $E_F^N$ for $z>d$
(this allows one to formulate quasiclassical equations in the region
$|z|>d$). Here we treat the contact between the $S$ and $N$ in a
fully microscopic fashion: starting from the
Hamiltonian~(\ref{eq:tbh}), a Fermi level mismatch $\Delta E_F =
E_F^N - E_F^S$, and assuming a screening length $l_D$ of a few
lattice spacings, we find the charge redistribution around the
contact, as well as the corresponding classical electrostatic
potential generated by them. Thus, our technique can treat
arbitrary spatial variation of the (lattice) Green functions,
superconducting order parameter, and electrostatic potential. This
includes the region $|z|<d$, where we find a sharp increase of
$V(z)$  but never as sharp as the (unphysical) delta function.

Since our $SINIS$ multilayer structure is translationally
invariant in the transverse direction,  each infinite plane has a
uniform surface charge distribution $\delta n(z) a$ which
generates a homogeneous electric field $E(z)=\delta n(z)
a/2\epsilon_0\epsilon_{\infty}$ pointing along the $z$ direction
($\epsilon_\infty$ is the relative dielectric constant of the
ionic lattice). The quantum-mechanical part of the electrostatic
problem entails determining the local electron density $n(z_i) \equiv n(z)$
(filling) at each site of a given plane $\alpha=z/a$
\begin{equation}\label{eq:filling}
n(z) = k_{B}T\sum_{\omega_{n}}
\int\limits^{\infty}_{-\infty}\rho^{2D}(\varepsilon_{xy}) \,
\mbox{\large Im} \, G_{\alpha\alpha}(i\omega_{n},
\varepsilon_{xy}) \, d\varepsilon_{xy},
\end{equation}
where $\varepsilon_{xy} = -2t\left[ \cos(k_{x}a) + \cos
(k_{y}a)\right]$ is the in-plane kinetic energy for the
transverse momentum $(k_x,k_y)$, and $\rho^{2D}(\varepsilon_{xy})$
is the two-dimensional tight-binding density of states on a
square lattice (which is used for the sum over momenta parallel
to the planes). The corresponding electric potential is
determined classically from the ``charge
deviation'' $\delta n(z)=n(z)-n$ ($n$ is the average filling
in the bulk, $n_N$ or $n_S$)
\begin{equation}\label{eq:deltav}
\delta V(z) = -\frac {e a\delta n(z^\prime)|z-z^\prime|}{2\epsilon_0\epsilon_\infty}.
\end{equation}
This must be summed over all planes to give the on-site potential
$V(z)$. Therefore,  the small induced charge imbalance $\delta
n(z) = N(\mu) e \delta V(z)$ satisfies (in a corresponding
continuous system)
\begin{equation}
\frac{d}{dz}\delta n(z)=-\frac{e^2 a N(\mu)}{2\epsilon_0 \epsilon_{\infty}}
\delta n(z)
\end{equation}
where $N(\mu)$ is the total density of states at the chemical potential
$\mu$. This is integrated to give the distribution of the
screened charge
\begin{equation}
\delta n(z)=\delta n(z_0) \exp \left
[ -\frac{e^2 a N(\mu)}{2\epsilon_0 \epsilon_{\infty}} (z-z_0) \right],
\end{equation}
which decays exponentially on a length scale set by the Debye
screening length
\begin{equation}
l_D=\left[ \frac{e^2 a  N(\mu)}{2\epsilon_0\epsilon_{\infty}} \right]^{-1}.
\end{equation}
Thus, the screening length is determined by $N(\mu)$ and
$\epsilon_\infty$ (for example,~\cite{gurevich98}
$\epsilon_\infty=20-30$ and $l_D=5-10 \, {\text \AA}$ in
high-$T_c$ superconductors). We choose
$\epsilon_\infty^S=\epsilon_\infty^N=5.0$ which leads to $\l_D \approx 3a$.
The self-consistency in the electrostatic problem is required because
$V(z)$ enters into the
computation of the Green function as a diagonal potential in the
Hamiltonian~(\ref{eq:tbh}). The solution has converged when the
potential is consistent with the charge
distribution~(\ref{eq:filling}) determined from the Green
function. Although this seems like a cumbersome computational
task, the potential around the $SN$ boundary barely changes when
equilibrium Josephson current flows. Thus, the electrostatic part
of the  problem converges rapidly since the potential found in the
solution at one phase gradient is a good initial guess for the
iteration scheme at the next superconducting phase gradient.

The density of electrons $n(z)$ on each site in a given plane (at zero
supercurrent) is plotted as a function of the junction thickness
for $\Delta E_F=3.0$ in Fig.~\ref{fig:n_def3}. The charge deviation $\delta
n(z)$ from half-filling $n_S=n_N=1$ and the corresponding
electrostatic potentials are (approximately) symmetric around the $SN$ boundary
for thick enough junctions, as shown in Fig.~\ref{fig:dn_L20}.
Strictly speaking, only such symmetric distributions should be
denoted ``screened dipole layers''. For thinner junctions, where
the screening of the excess charge does not heal $n(z)$ to its
equilibrium value, charge is depleted from the $N$ interlayer.
The example of this behavior is the $L=2a$ junction in the lower panel
of Fig.~\ref{fig:dn_L20}. It leads to a non-monotonic resistance
as a function of junction thickness $L$ at fixed $\Delta E_F$
(see Sec.~\ref{sec:icrn}). Thus, the charge effects become
essential in short-coherence length superconducting junctions
with thicknesses $L < 2l_D$, which are encountered in high-$T_c$
grain boundaries.~\cite{gurevich98}

The interesting feature of the $\delta n(z)$ profiles for the
half-filled $S$ and $N$ is that they can be approximately
rescaled to a single reference distribution [set by $(\Delta
E_F)_{\rm ref}$] after multiplying each of them by the ratio
$(\Delta E_F)_{\rm ref}/\Delta E_F$, as shown in the insets of
Fig.~\ref{fig:dn_L20}. We believe this occurs because the
noninteracting cubic density of states is nearly constant close
to half filling. The scaling becomes essential in computing the
properties of junctions with large $\Delta E_F$ since one can use
the scaled potential profile computed at a smaller $\Delta E_F$
as the initial guess in the iteration procedure. Since the
$n_S=n_N=1$ case has a higher degree of symmetry, we also perform
calculations for $n_N=0.01$ (which approximates a doped
semiconductor) in the normal region and half-filling $n_S=1.0$ in
the superconductor. The result is shown in
Fig.~\ref{fig:n001_def1}. Here the scaling of the $\delta n(z)$
distribution does not work as well (because the density of states has
strong variation with energy). In addition, we find that the
charge deviation is nonsymmetric, and yields a different  $\delta
n(z)$ for positive and negative $\Delta E_F$ [for symmetric
filling $n_S=n_N$ the two distributions are simply related as
$\delta n(z)|_{-\Delta E_F} = - \delta n(z)|_{\Delta E_F}$].  We
also investigate the temperature dependence of the distributions
of uncompensated charge for $T=0.2$ (the chemical potential in
the bulk $N$ is $\mu=-5.566$ for $n_N=0.01$) and at $T=0.09$
(which is close to $T_c=0.11$). In both cases $n_S=n_N=1.0$ and
$n_S \neq n_N=0.01$ we find $\delta n(z)$ to be practically
temperature independent (e.g., the change is at most $5 \%$
around the $SN$ boundary) for $\Delta E_F$ shown in the previous
figure. This feature is exploited in Sec.~\ref{sec:icrn} to
calculate the normal state resistance of our junctions from an
imaginary axis computation of the charge and potential profile in
the superconducting state. However, for $n_S \neq n_N=0.01$ and
small $|\Delta E_F| \simeq 0.2$ a large change in the magnitude of
$\delta n(z)$  is observed when going from $T=0.2 > T_c$ to 
$T=0.01 < T_c$, as shown in Fig.~\ref{fig:change_dn}. 
Similar phenomenon has been found in the recent Raman 
studies~\cite{igor} which show a substantial change in the 
thickness of the charge accumulation 
layer at the interface between Nb and InAs, as Nb undergoes 
a superconducting transition and proximity effects develop in 
the InAs layer. This would point to a proximity effect 
influenced screening length, which cannot be seen in our local 
(Thomas-Fermi) screening theory containing only 
two parameters which determine $l_D$: $\epsilon_\infty$, which is 
fixed in our calculations, and the density of states $N(\mu)$ 
which can be modified by the proximity effect. Our observation 
of the change in the charge concentration above ($T \sim \Delta E_F$) 
and below $T_c$, without a palpable change in the screening properties, 
suggest that effects beyond the simple screening theory 
(e.g., nonlocal screening which becomes important in low filling 
cases~\cite{gurevich98}) probably have to be taken into account to 
understand this experiment.

\section{Self-consistent equilibrium properties of $SINIS$ Junctions}
\label{sec:phase}

We first provide an insight into the microscopic properties of
these junctions which are determined by the proximity effect that
affects the critical current (in non-self-consistent calculations such
effects are taken into account only through some effective
phenomenological suppression parameter~\cite{sinis_review}). They
are encoded in the self-consistently computed variation of the
amplitude and phase of the order parameter
$\Delta(z)=|\Delta(z)|e^{i\phi(z)}$ in the $S$ or pair
amplitude $F(z)=|F(z)|e^{i\phi(z)}$ in the $N$. These are
related to each other inside the $S$ by
\begin{equation}\label{eq:delta_f}
  \Delta(z)=-U(z)F(z).
\end{equation}
where $F(z)$ is obtained as the equal-time limit of the local
anomalous Green function introduced in Sec.~\ref{sec:lattice}
\begin{equation}\label{eq:f_ii}
  F(z)=F(z_i,z_i,\tau=0^+).
\end{equation}
Although $\phi(z)$ and $F(z)$ are not directly measurable, they
are important for understanding the superconductivity in
inhomogeneous structures. Examples include  the proximity effect in
the $N$ and the depression of $|\Delta(z)|$ (compared to its bulk
value) on the $S$ side of a $SN$ boundary (``inverse proximity
effect''). Since the critical current of the junction is
determined by $\Delta(z)$ at the $SN$ boundary, the study of
$F(z)$ throughout the junction gives direct insight into how
self-consistency affects the transport properties (analytical
approaches usually assume a step function for $|\Delta(z)|$, which
is applicable only for a limited range of junction
parameters~\cite{likharev_review}). The non-zero value of $F(z)$
inside the superconductor results from the attractive pairing
interaction $U(z) \neq 0$ [which also gives rise to the non-zero order
parameter $\Delta(z)$]. In the normal metal, $U(z)=0$ and the gap
vanishes, but $F(z)$ can be non-zero due to the proximity effect.
Therefore, it is more meaningful to plot $F(z)$, which is a
continuous function throughout the junction. Inside the $S$,
$F(z)$ should be understood as just $\Delta(z)/[-U(z)]$. The
superconducting 
\begin{figure}
\centerline{\psfig{file=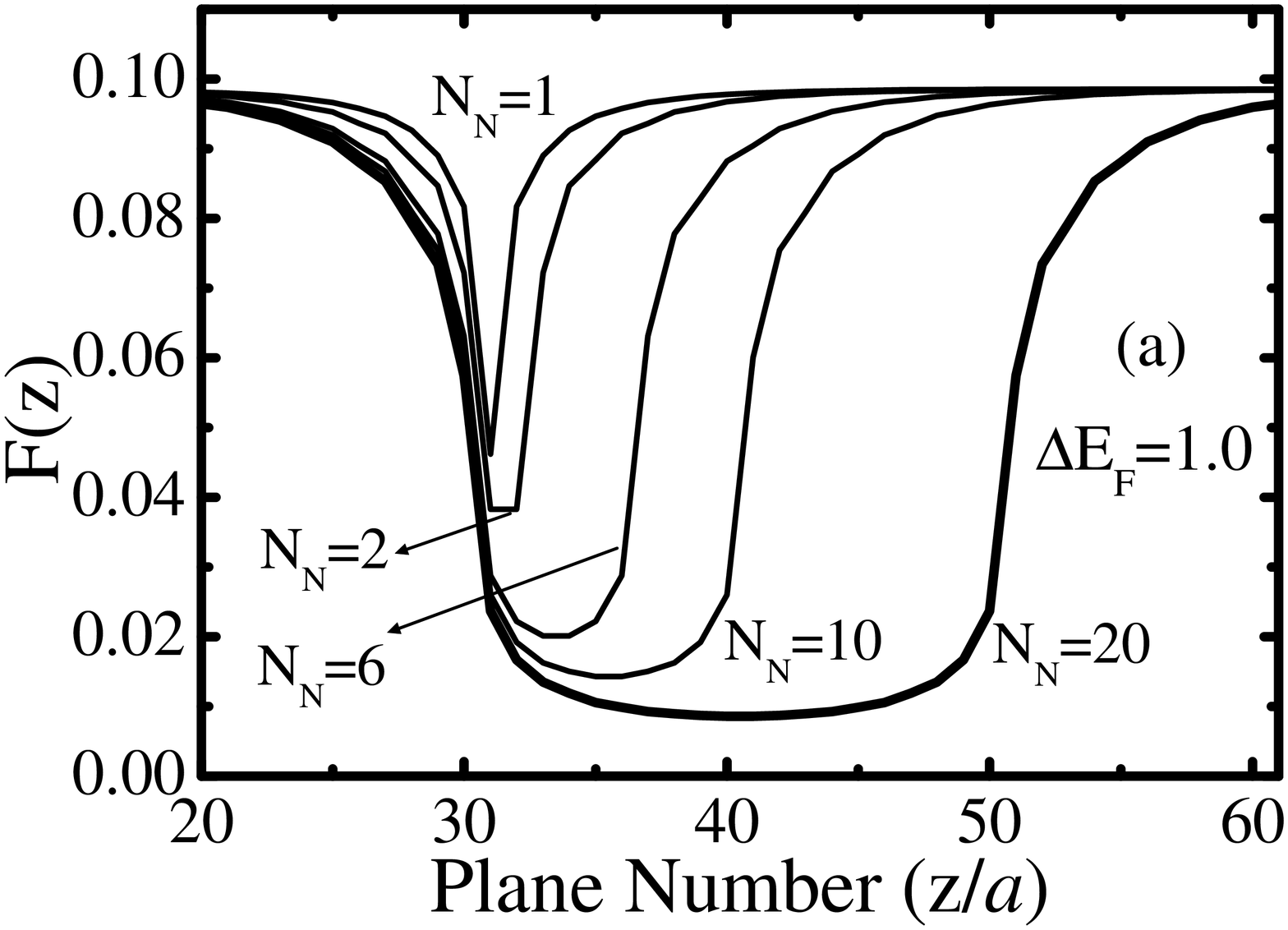,height=3.0in,angle=-90} }
\vspace{0.2in}
\centerline{\psfig{file=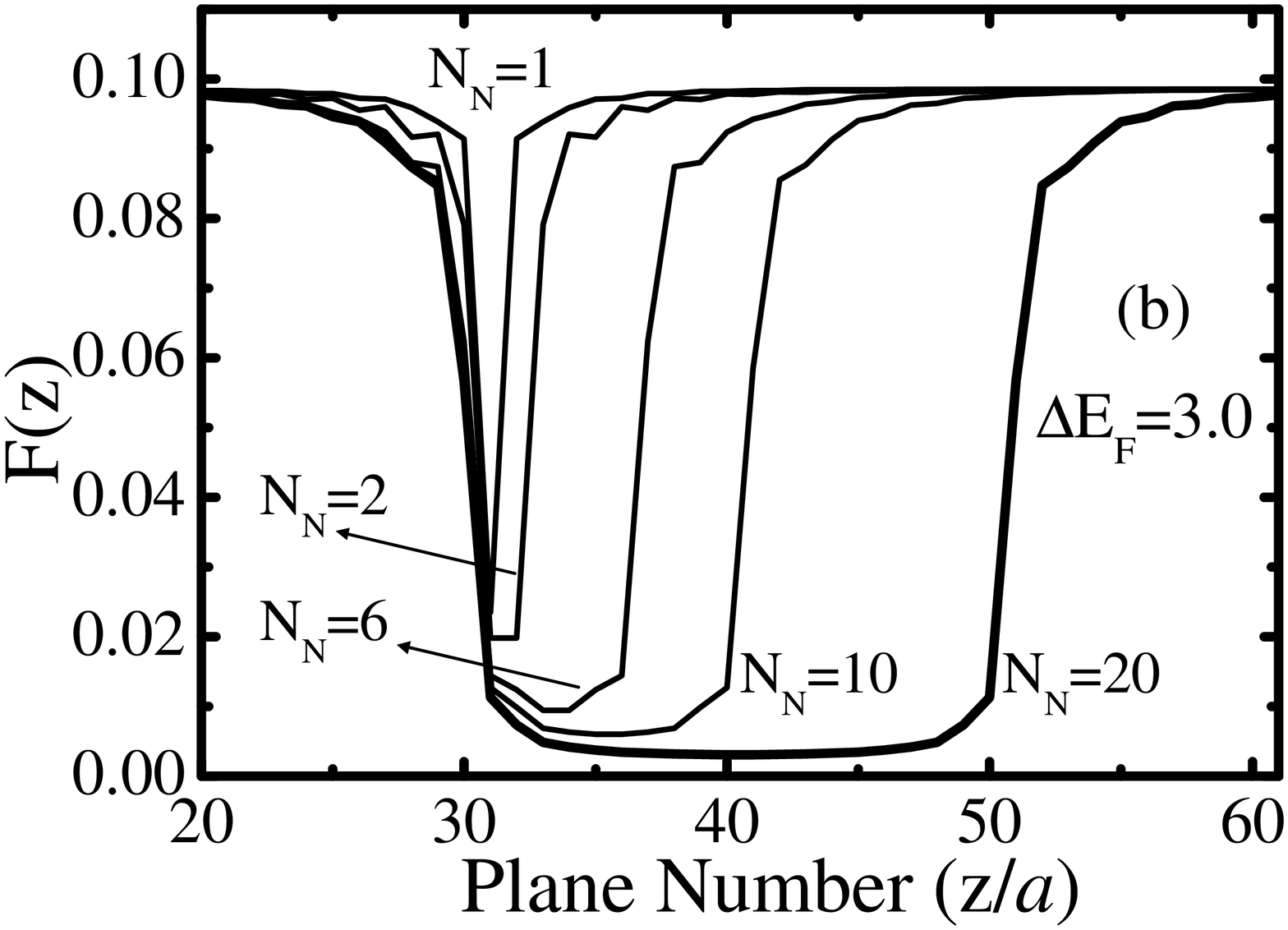,height=3.0in,angle=-90} }
\vspace{0.2in} \caption{Pair amplitude $F(z)$ at zero supercurrent
flow in $SINIS$ junctions of different thicknesses $L=N_Na$. The
double-barrier structure arises from the charge inhomogeneity
(see Fig.~\ref{fig:n_def3}) induced by the difference between the
Fermi energies of the normal metal and superconductor: (a)
$\Delta E_F = E_F^N-E_F^S =1.0$, and (b) $\Delta E_F =
E_F^N-E_F^S =3.0$.} \label{fig:F_def1}
\end{figure}
correlations are imparted to the $N$ region which
is in contact with the $S$. They are described quantitatively 
by the pair amplitude~\cite{deutscher} $F(z)$~(\ref{eq:f_ii}).
Because of the translational symmetry of the junction in the
transverse direction, $F(z)$ is constant within the plane, and
changes from plane to plane along $z$-axis. The scale over which
$F(z)$ changes exponentially from the $SN$ interface to zero in
the bulk of the $N$ is set by the normal metal coherence length
$\xi_N$. However, as $T \rightarrow 0$ the length $\xi_N$ diverges
and the exponential decay of $F(z)$ crosses over to a slower
power-law
decay (like $1/z$ at $T=0$, inside a $N$ described by a Fermi
liquid~\cite{falk}).

Although this description of the proximity effect has
been used
since the early days of inhomogeneous superconductivity
studies,~\cite{deutscher} it is only recently that mesoscopic
superconductivity~\cite{superlattices} has established an explicit
connection to a real-space picture of pairing correlations,
provided by the phenomenon of (phase-coherent) Andreev
reflection.~\cite{andreev} That $F(z)$ is non-zero in a normal
region is equivalent to saying that the electron and an Andreev
reflected phase-conjugated hole maintain their single-particle
phase coherence inside the $N$. Technically, this interpretation
follows directly from the expression for $F(z)$ in terms of
quasiparticle wave functions entering the BdG
equations.~\cite{carlo_rmt} In other words, near the $SN$
boundary, Andreev reflection mixes electron-like and hole-like
quasiparticles in the same proportion in which they are mixed in
the $S$ (where Bogoliubov quasiparticles are a mixture of
electron-like and hole-like states with weights determined by the
self-consistency condition) due to purely kinematic effects,
since the interaction is absent in the $N$. The definition of
$F(z)$ from Eqs.~(\ref{eq:g_nambu}) and ~(\ref{eq:f_ii}) in the
second-quantized formalism, shows that such correlations can be
interpreted alternatively as a condensate wave function leaking
into the normal metal through the presence of evanescent Cooper
pairs.~\cite{lesovik} In the case of a Josephson junction, the 
overlap of two condensate wave
functions provides a weak coupling
between the superconducting leads, while insuring the global
phase coherence and equilibrium current flow (i.e., the DC
Josephson effect) for the time-independent phase difference
between them. Thus, the two apparently different pictures of the
Josephson effect in weak links (leakage of Cooper pairs versus
Andreev reflection induced transfer of Cooper pairs) are in fact
two facets of the same phenomenon.

\begin{figure}
\centerline{\psfig{file=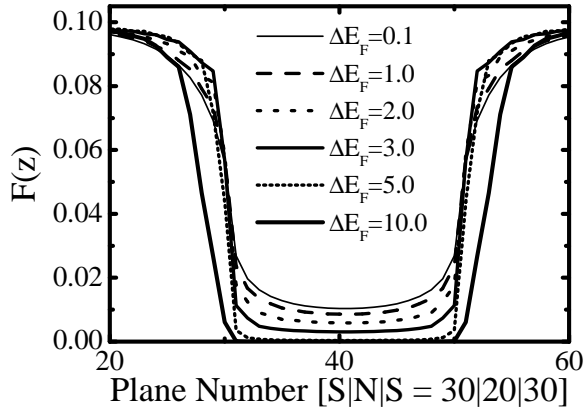,height=3.0in,angle=-90} }
\vspace{0.2in} \caption{Pair amplitude $F(z)$ at zero supercurrent
flow in $SINIS$ junctions of thickness $L=20 a$ characterized by
different heights of the SDL barriers. The double-barrier
structure (charge inhomogeneity from Fig.~\ref{fig:dn_L20})
arises from the difference in Fermi energies of the normal metal
and superconductor, $\Delta E_F = E_F^N-E_F^S$.} \label{fig:F_L20}
\end{figure}

\begin{figure}
\centerline{\psfig{file=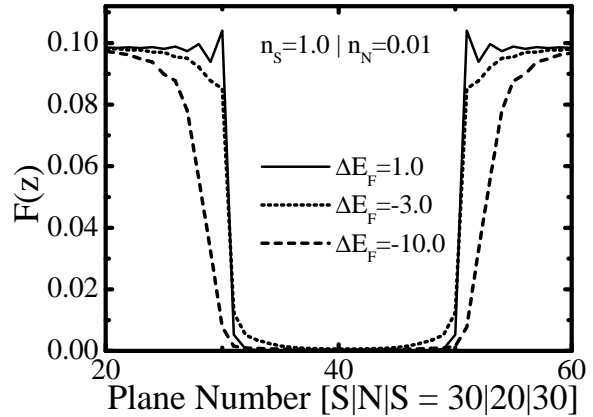,height=3.0in,angle=-90} }
\vspace{0.2in} \caption{Pair amplitude $F(z)$  at zero
supercurrent flow in $SINIS$ junctions with thickness $L=20a$.
The double-barrier structure arises from the inhomogeneous charge
redistributions, plotted in Fig.~\ref{fig:n001_def1}, induced by
$\Delta E_F = E_F^N-E_F^S$ between the normal region with
$n_N=0.01$ and the superconductor at half-filling $n_S=1$.}
\label{fig:F_def5}
\end{figure}

We first show two examples of $F(z)$ computed self-consistently
for vanishing supercurrent throughout the
$SINIS$ junction with $n_S=n_N=1$. Figure~\ref{fig:F_def1} plots the scaling
of $F(z)$ with the junction thickness for the $I$ layers at the $SN$ boundary being
SDLs whose height is determined by $\Delta E_F=1.0$ or $\Delta
E_F=3.0$. The shape of $F(z)$ evolves with the thickness, as well
as with the height of the double-barrier. This second point is
demonstrated in Fig.~\ref{fig:F_L20} where we fix $L$ and vary
the strength of the SDL barrier. Here one would expect the
evolution of $F(z)$ toward a step function, which then justifies
the use of rigid boundary conditions for strong enough scattering
at the $I$ barriers.~\cite{likharev_review} However, we find a
non-monotonic change in the shape of $F(z)$: the influence of a
SDL on the order parameter $\Delta(z)$ is first reduced with
increasing $\Delta E_F$, but then leads to a depressed
$\Delta(z)$ near the boundary for a strong charge imbalance
generated by $\Delta E_F=10.0$. Since our previous results for a
$SINIS$ junction having a strong on-site Coulomb potential,
confined within a single plane, exhibit a step function
like~\cite{miller} $\Delta(z)$, the effects observed here can be
attributed to the finite spatial extent of the SDL. Moreover, we
find that the step function (up to tiny oscillations near the
boundary) for $\Delta(z)$ does develop in the special case of low
filling in the $N$ region, like $n_N=0.01$, and a small mismatch
$|\Delta E_F| \lesssim 1$. A specific example of this behavior
(compared to the case with the same parameters, but with a negative
$\Delta E_F$) is shown in Fig.~\ref{fig:F_def5}.

\begin{figure}
\centerline{\psfig{file=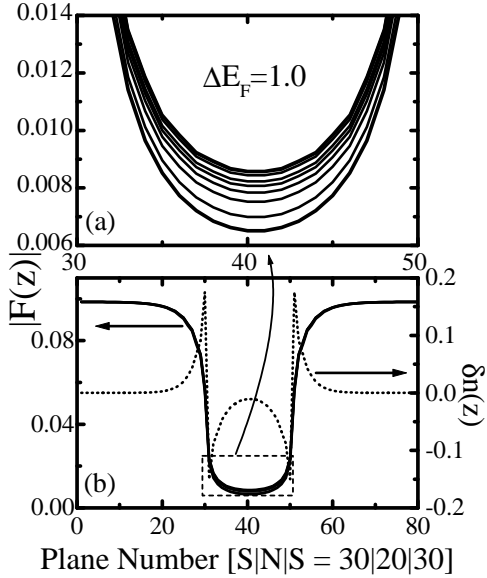,height=3.0in,angle=0} }
\vspace{0.2in} \caption{Pair amplitude $|F(z)|$ at different
supercurrent flows through the $SINIS$ junction as a response to
a non-zero phase gradient in the bulk and corresponding phase
change $\phi$ across the $N$ region. The thickness of the
junction is $L=20 a$, and the two SDLs are generated by the Fermi
energy mismatch $\Delta E_F=1.0$ between the $N$ and $S$, which are
both at half filling. Panel (a) is a blow up of the dashed square
region in panel (b).} \label{fig:F_ic}
\end{figure}

In the short junction case, the oscillations of $\Delta$ on the
scale of $\lambda_F$ are observed for large enough $\Delta E_F$.
In this case, as discussed in the previous section, the junction
is too thin for the distribution of charge to heal to its
equilibrium value. The charge depletion inside the $N$
brings it close to an insulating state. While oscillations 
on the scale of
$\lambda_F$ have been observed~\cite{levy} in similar
self-consistent calculations at $T=0$ (and attributed to the
mesoscopic coherence of a single particle wave function), here it 
appears that they are a property of the superconducting interface
which terminates at an ``insulator'' (this is also exhibited by a
thick junction with small $n_N$ in Fig.~\ref{fig:F_def5}).
We have recently found such behavior, in its most pronounced form,
in the case of $SIS$ junctions, with $I$ being a correlated
insulator.~\cite{freericks} In the self-consistent calculations
one can also observe how $F(z)$ evolves, becoming smaller inside
the $N$ region, as the phase change across the interlayer is
increased and the Josephson current approaches $I_c$. An example
of such an effect due to self-consistency is shown in Fig.~\ref{fig:F_ic}.

\begin{figure}
\centerline{\psfig{file=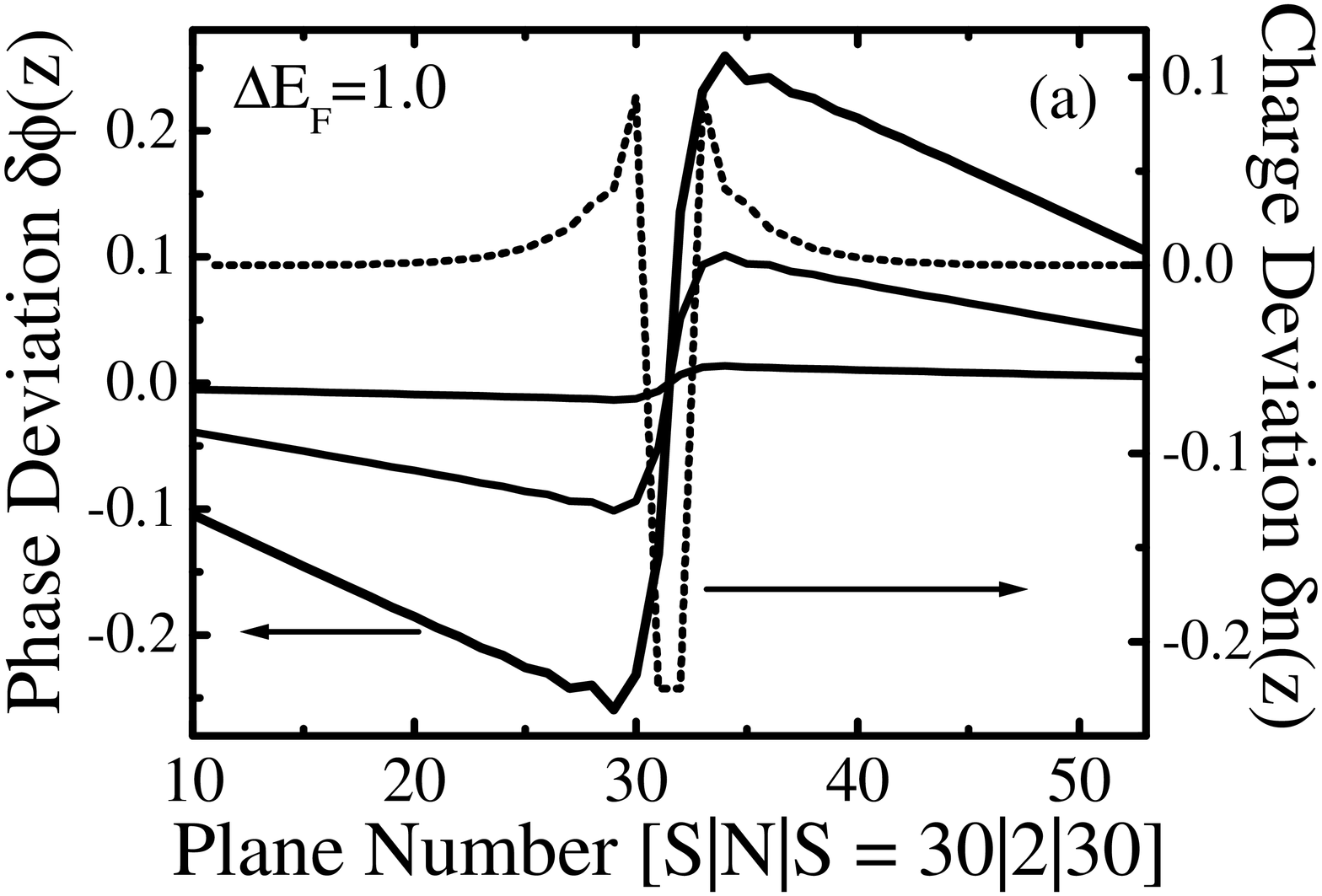,height=3.0in,angle=-90} }
\vspace{0.3in}
\centerline{\psfig{file=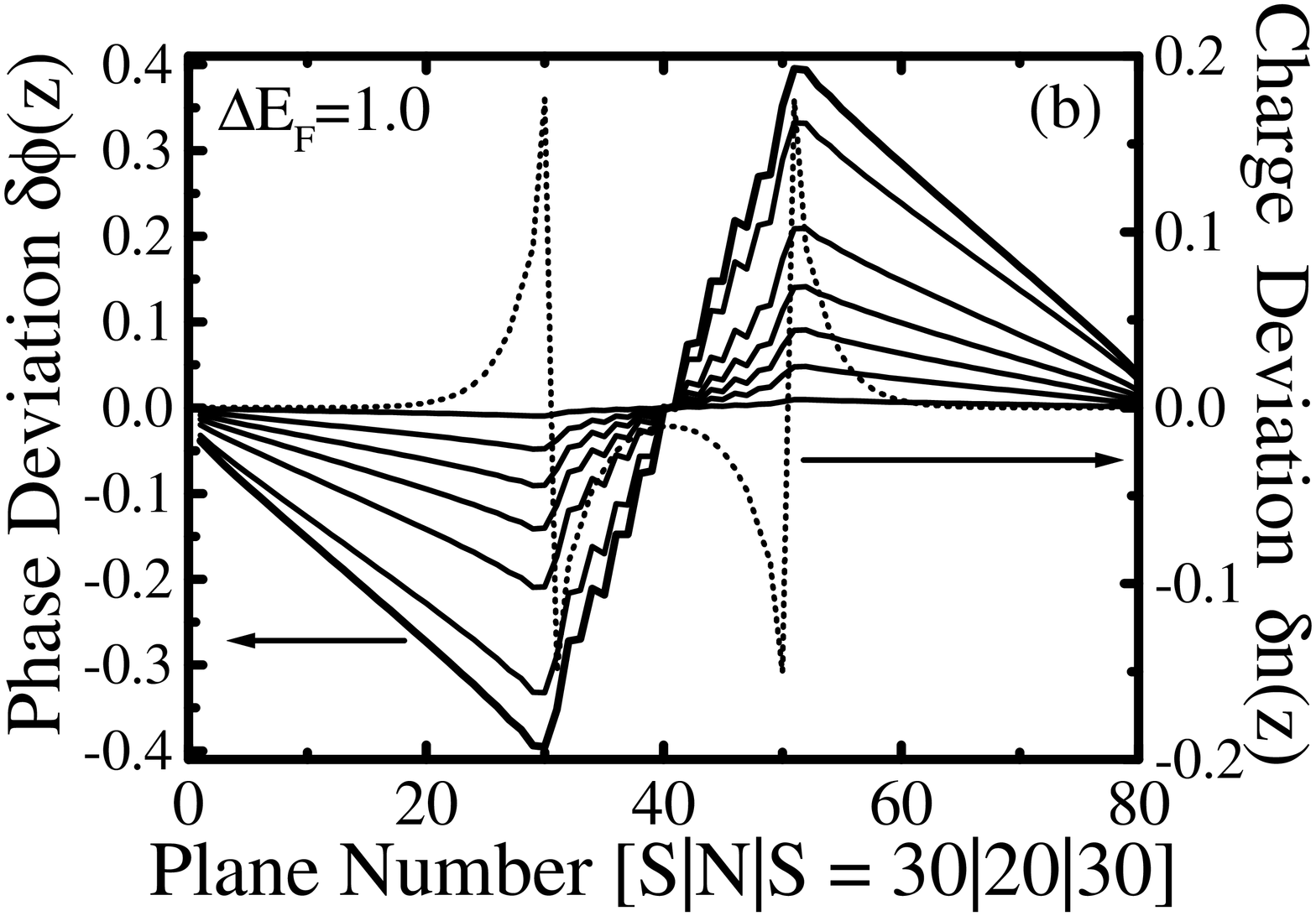,height=3.0in,angle=-90} }
\vspace{0.2in} \caption{Phase deviation $\delta \phi(z)$ [the
total phase change along $z$-axis is the sum of the bulk phase
gradient and $\delta \phi(z)$, Eq.~(\ref{eq:phase})], within the
self-consistently modeled part of the $SINIS$ junction, at
different supercurrent flows (the last curve corresponds to
$I_c$). The thicknesses of the junctions are: (a) $L=2 a$, and (b)
$L=20a$, and the double-SDL-barrier (Fig.~\ref{fig:dn_L20}) is
induced by the Fermi energy mismatch $\Delta E_F=1.0$ between the
$N$ and $S$.} \label{fig:phaseL20}
\end{figure}

When self-consistency is satisfied, the phase of the order
parameter $|\Delta(z)| e^{i\phi(z)}$ is not a constant inside the $S$
leads (see also Sec.~\ref{sec:icrn}) because a phase gradient is
needed to support the current in the $S$ ensuring current
conservation throughout the junction. Thus, the change of phase
from plane to plane has to be extracted
from the self-consistent solution for $|F(z)|e^{i\phi(z)}$. It can
be expressed as the sum of a linear term and a ``phase
deviation'' term $\delta \phi(z)$
\begin{equation}\label{eq:phase}
  \phi(z) = z \left( \frac{d\phi}{dz}
  \right)_{\rm bulk} + \delta \phi(z),
\end{equation}
where the distance $z$ is measured from the origin along the
$z$-axis. The linear term is determined by the phase
gradient $(d \phi/dz)_{\rm bulk}$ which is set as the boundary condition in the
bulk of the superconductor. The non-trivial information contained
in $\phi(z)$ is revealed by plotting $\delta \phi(z)$. The
overall phase $\phi(z)$ increases smoothly and monotonically across
the self-consistent region.  We plot 
$\delta \phi(z)$  for two different
junction thicknesses with
$\Delta E_F=1$ in Fig.~\ref{fig:phaseL20}. In general,
oscillations of $\delta \phi(z)$ on the scale of $\lambda_F$ are
found for moderate $\Delta E_F$ and long enough junctions ($L
\gtrsim 6a$). Oscillations, of both the phase and the order
parameter, were found inside a long mesoscopic constriction in
previous self-consistent calculations~\cite{levy}  at zero
temperature (that gradually disappear with increasing $T$). Here
we see the oscillations of $\delta \phi(z)$ at low temperature
(but still $L < \xi_N$), while the corresponding $|F(z)|$
(Fig.~\ref{fig:F_def1}) does not oscillate.

\section{Critical currents and characteristic voltages}
\label{sec:icrn}

In the self-consistent treatment, equilibrium supercurrent flows
through the junction when the phase gradient $(d \phi/dz)_{\rm
bulk}$ exists in the bulk of the superconductor and a total phase
change $\phi$ is established across the normal region. Therefore,
we first find the solution for the bulk superconductor in both the
absence of a supercurrent and in the presence of a supercurrent
generated by a uniform variation in the order-parameter phase.
The uniform bulk solution is then employed to provide the
``boundary conditions'' for the junction beyond the region where
we determine properties self-consistently. Thus, our method does
not require any assumptions about the boundary conditions at the
interface between the barrier and the superconductor, which
follow from the requirements of self-consistency.~\cite{miller}
We use current conservation as a stringent test of the achieved
self-consistency in the solution for the Green function. Namely,
the self-consistently determined $\Delta(z)$ ensures that
Andreev reflection at each $SN$ boundary generates supercurrent
flow in the $S$ leads (besides being responsible for the
proximity effect in 
the $N$ discussed in Sec.~\ref{sec:lattice}).
Thus, the fulfillment of the self-consistency condition~(\ref{eq:delta_f}) 
means that the ``source term'' (on the right-hand side) vanishes in the equation of 
motion for the charge density operator $\hat{n}_i$
\begin{equation} \label{eq:conservec}
\frac{\partial \hat{n}_i}{\partial t} + \sum_j I_{ij} = \frac{2ie}{\hbar} \left(
\Delta_i \langle c_{i\uparrow}c_{i\downarrow} \rangle -
\Delta^*_i \langle c_{i \downarrow}^\dag
c_{i \uparrow}^\dag \rangle \right ),
\end{equation}
thereby recovering current continuity at every site ($I_{ij}$ is
the current between two neighboring sites). When the current inside
the superconductors is small,  e.g., due to the geometrical
dilution of a weak link with a junction area much smaller than
$\xi_0^2$, or when the junction resistance is dominated by a large
interlayer resistance,~\cite{carlo_rmt,likharev_review} one
usually neglects the supercurrent flow and corresponding phase
gradient in the bulk superconductor necessary to support it.
Strictly speaking, such approaches violate current
conservation.~\cite{sols,bagwell} Inasmuch as our $S$ and $N$
layers have the same area, $I_c/I_c^{\rm bulk}$ can be close to
one for thin junctions with weak SDLs at small $\Delta E_F$. In
such cases, current flow affects appreciably the superconducting
order parameter [i.e., $F(z)$ both inside and outside of the $N$,
cf. Sec.~\ref{sec:phase}] and a self-consistent treatment becomes
necessary (as is the case for the critical current of the bulk
superconductor~\cite{bardeen}). Because of the presence of a
phase gradient inside the $S$, the simple picture~\cite{josephson}
of an equilibrium current being
related to the phase difference $\phi_L-\phi_R$ between the left
and right $S$ leads (where $\phi_L$ and $\phi_R$ are constant
within the leads) is not applicable. Nevertheless, the solution
for the current turns out 
to be uniquely parameterized by a
single quantity which can be taken as the phase change across the
$N$ region~\cite{kupriyanov92} (the other option is the phase
offset~\cite{sols} which is related to the phase change by a
nontrivial scale transformation). In a discrete model like ours, a
convention has to be introduced on how this change is extracted
from $\phi(z)$ in Eq.~(\ref{eq:phase}). The thickness of the
junction is defined to be the distance measured from the point
$z_L$, in the middle of the last $S$ plane on the left (at
$z_L^S$) and the first adjacent $N$ plane (at $z_L^N=z_L^S+1$), to the
middle point $z_R$ between the last $N$ and first $S$ plane on
the right (cf. Fig.~\ref{fig:planes}). Since $\phi(z)$ is defined
within the planes, we set $\phi(z_L)=[\phi(z_L^S)+\phi(z_L^N)]/2$
to be the phase at $z_L$, and equivalently for $\phi(z_R)$. The phase
change across the barrier is then given by
\begin{equation}\label{eq:phasechange}
  \phi = \phi(z_R)-\phi(z_L) = L\left( \frac{d\phi}{dz}
  \right)_{\rm bulk} + \delta \phi(z_R)- \delta \phi(z_L),
\end{equation}
\begin{figure}
\centerline{\psfig{file=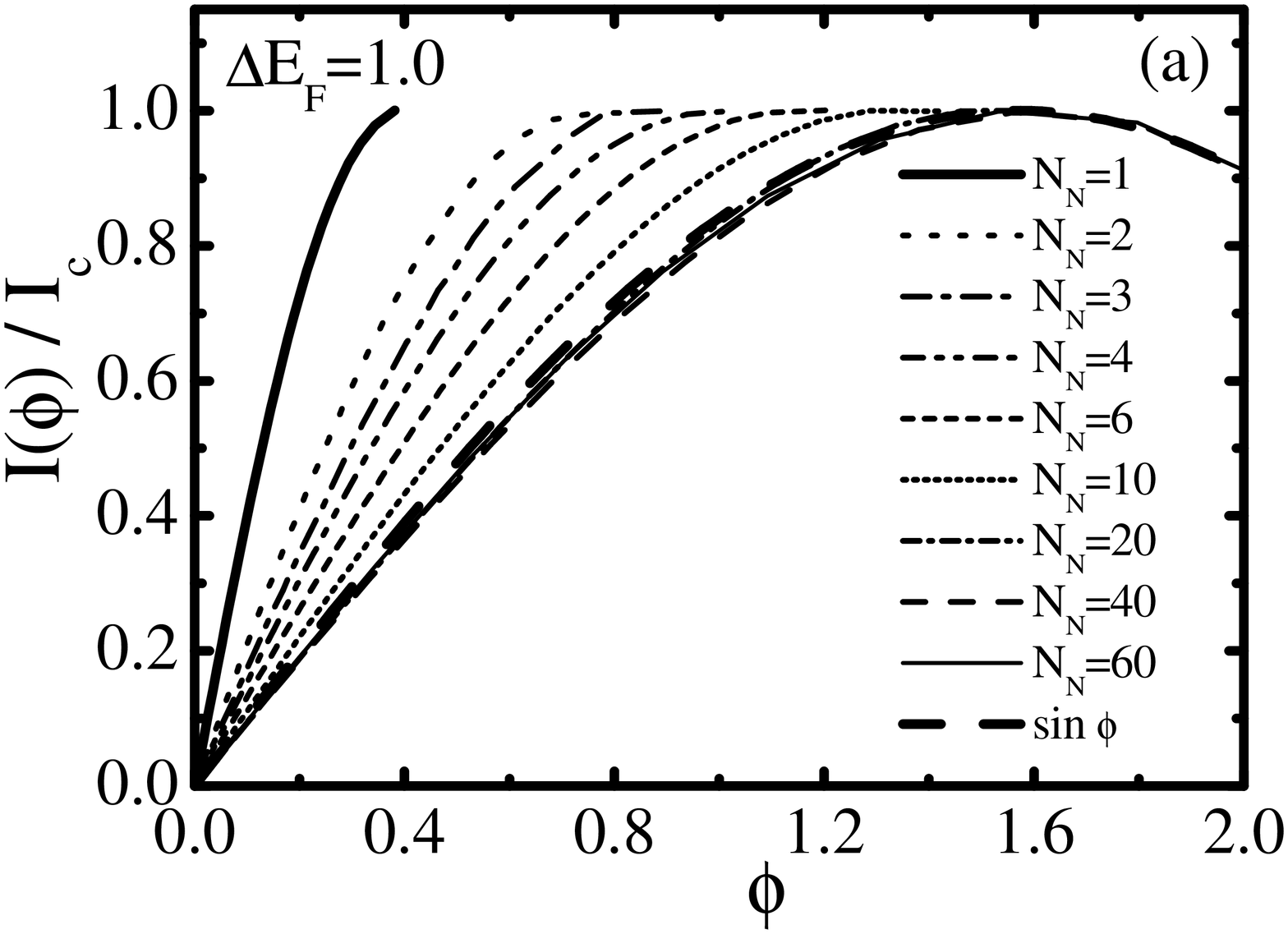,height=2.6in,angle=-90} }
\vspace{0.3in}
\centerline{\psfig{file=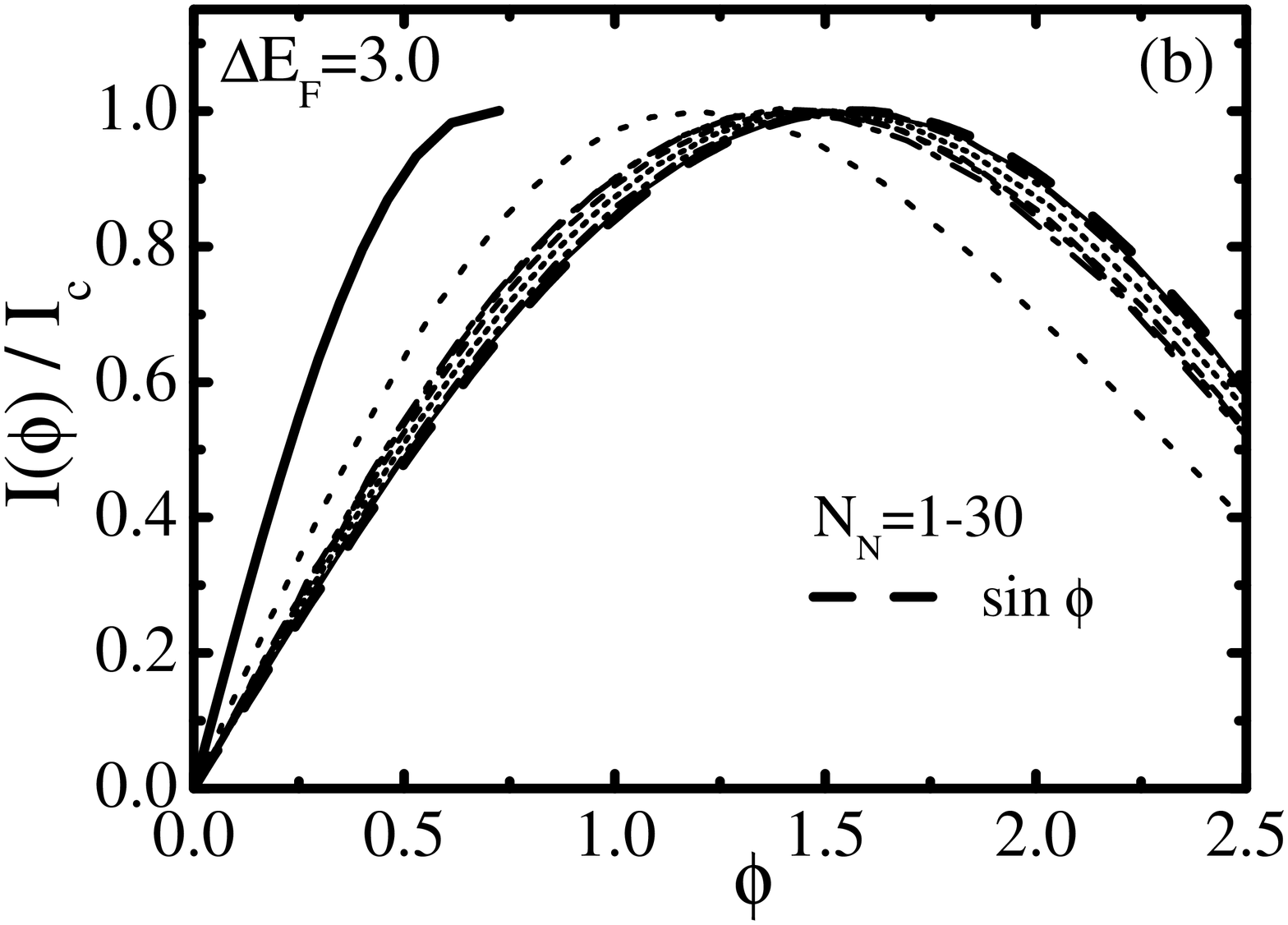,height=2.6in,angle=-90} }
\vspace{0.2in} \caption{Scaling of the current-phase relation
$I(\phi)/I_c$ with the thickness of the $SINIS$ junctions where
the SDLs are determined by: (a) $\Delta E_F=1.0$, and (b) $\Delta E_F=3.0$.
Note that the phase change $\phi_c$ at the critical current $I_c=I(\phi_c)$
varies nonmonotonically with the junction thickness, as shown in
Fig.~\ref{fig:phi_c}. The standard  $I(\phi)/I_c=\sin \phi$
dependence in the $SIS$ tunnel junction~\cite{josephson} is plotted
as a reference only (which is analytically predicted for
$SINIS$ junctions with small barrier transparency at high
enough temperatures~\cite{sinis_review}).}
\label{fig:cpc}
\end{figure}
for a junction of thickness $L$. The current-phase relation $I(\phi)$ 
is obtained by computing the current for a fixed bulk phase gradient 
and associating this value with the phase change across the $N$ region, 
which is extracted from the self-consistent pair amplitude $F(z)=|F(z)|e^{i
\phi(z)}$. On the lattice, transport is described by the current
across a link between two adjacent planes $\alpha$ and
$\alpha+1$. This current (per $a^2$) is obtained from the Green
function connecting two neighboring planes as~\cite{miller}
\begin{eqnarray}\label{eq:current}
I_{\alpha,\alpha+1} & = &
\frac{2eat}{\hbar}k_{B}T\sum_{\omega_{n}}
\int\limits^{\infty}_{-\infty}\rho^{2D}(\varepsilon_{xy})
\nonumber \\
&& \times {\rm Im} \left[ G_{\alpha,\alpha+1}(i\omega_{n},
{\varepsilon_{xy}}) \right] d \varepsilon_{xy}.
\end{eqnarray}
The first iteration in our self-consistent algorithm usually gives
a current which is smaller inside the $N$ than in the $S$ region.
The iteration cycle is completed when the current is constant
throughout the junction. The only approximation invoked here is
the presence of a (typically small) discontinuity in the
supercurrent at the 
\begin{figure}
\centerline{\psfig{file=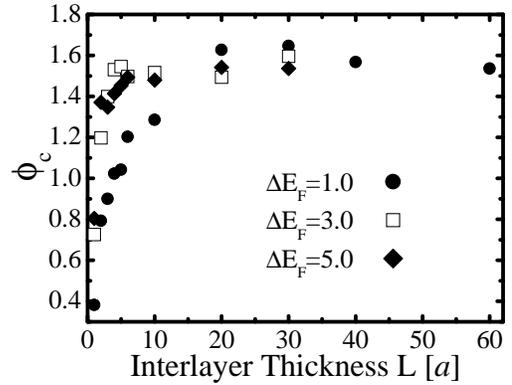,height=2.6in,angle=-90} }
\vspace{0.2in} \caption{Phase change $\phi_c$ at the critical current
$I(\phi_c)=I_c$ plotted against the thickness of the $SINIS$
junctions, where the double-SDL-barrier is
determined by $\Delta E_F=1.0$ (circles), $\Delta E_F=3.0$ (squares), and
$\Delta E_F=5.0$ (diamonds).}
\label{fig:phi_c}
\end{figure}
bulk-superconductor/self-consistent-superconductor interface. 
We find that the superconducting order has always healed to its bulk
value at that point. However, sometimes there can be a jump in
$\phi(z)$ at this boundary when one nears the critical current.
This discontinuity in the phase corresponds to a breakdown of
current conservation at the this interface (it can become large
for large $\Delta E_F$ and a thick junction, especially when one
lies on the decreasing current side of the current-phase
diagram). The critical current $I_c$ of the junction is reached
when the planes with the lowest pair amplitude $|F(z)|$ (which are
located in the center of the $N$) can no longer support the
necessary phase gradient to maintain current continuity.

\begin{figure}
\centerline{\psfig{file=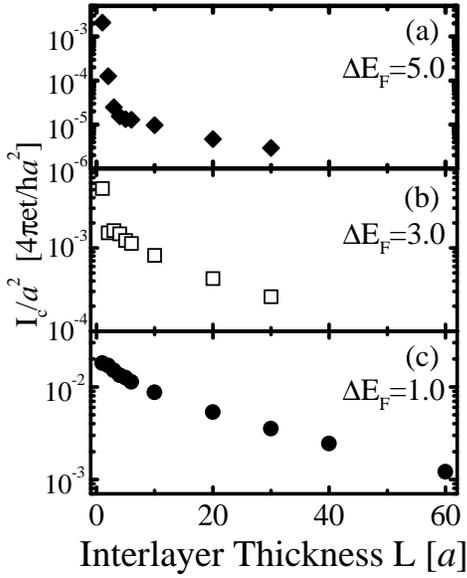,height=3.0in,angle=0} }
\vspace{0.2in} \caption{Semilogarithmic plot of the critical
current (per unit area $a^2$) as
a function of junction thickness (i.e., number of planes
$N_N=$1--60 inside the $N$ interlayer) of the $SINIS$ junction for
different SDLs determined by $\Delta E_F=E_F^N - E_F^S$. Both the
$S$ and $N$ are at half-filling in the bulk.} \label{fig:Ic_vs_L}
\end{figure}

\begin{figure}
\centerline{\psfig{file=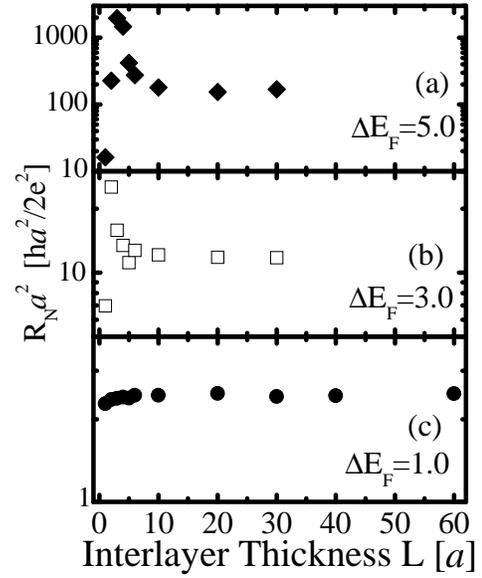,height=3.0in,angle=0} }
\vspace{0.2in} \caption{Semilogarithmic plot of the
normal state resistance (multiplied by the
unit area $a^2$), as a function of the $SINIS$ junction
thickness, for different SDLs determined by $\Delta E_F=E_F^N -
E_F^S$. Both the $S$ and $N$ are at half-filling in the bulk. Note
that junctions with thickness smaller than
$2l_D$ have charge depleted for large enough $\Delta E_F$
causing the resistance to change non-monotonically as a function
of $L$. The Sharvin point contact resistance of the clean $SNS$
junction (corresponding to $\Delta E_F=0$ in our structure) is
$R_{\rm Sh}a^2=1.58 ha^2/2e^2$} \label{fig:RN}
\end{figure}

The scaling of the shape of the current-phase relation with the
junction thickness is plotted in Fig.~\ref{fig:cpc} for different
$\Delta E_F$. We find large deviations from the usual sinusoidal $I(\phi)$
dependence for thin junctions and moderate heights of the SDL barriers.
While in such cases (and at low temperatures)
analytical predictions~\cite{brinkman,sinis_review,lukichev} also give
non-sinusoidal $I(\phi)$, our ``critical'' phase change $\phi_c$
[$I(\phi_c)=I_c$] is always below the analytical prediction
$\phi_c \approx 1.86$ (Fig.~\ref{fig:phi_c}), which can be attributed to
the effects of self-consistency~\cite{sols} (the other important distinction
is that SDLs are spatially extended barriers). For thicker junctions, with
high SDL barriers (and at high enough temperatures)
the recovery of the usual $SIS$ junction $I(\phi)=I_c\sin \phi$
current-phase relation is predicted.~\cite{sinis_review} Here we find a
current-phase relation $I(\phi)$ which is close to sinusoidal in
the thick junction limit [(a) panel in Fig.~\ref{fig:cpc}], or in thin 
junctions with high SDL barriers [(b) panel in Fig.~\ref{fig:cpc}].
The corresponding critical current densities as a function of junction thickness are
plotted in Fig.~\ref{fig:Ic_vs_L}. For large $\Delta E_F$,
$I_c(L)$ is non-monotonic because of the special role played by
the barriers formed in the junctions with $L < 2 \l_D$.
When SDLs are completely screened inside the thick $N$ interlayers,
the decay of current is determined just by the exponential decay of
the proximity 
coupling between the $S$ leads through the clean normal interlayer 
 (the resistance of these junctions is also practically independent
of $L$, see Fig.~\ref{fig:RN}). For example, the characteristic decay length,
extracted from fitting~\cite{likharev_review,jackson} $A L^p \exp[-L/\xi_N^\prime]$
to $I_c(L)$ for $\Delta E_F=1.0$ case in  Fig.~\ref{fig:Ic_vs_L},
is $\xi_N^\prime \approx 35a \simeq \xi_N$ (with $p=0.3$).
For larger $\Delta E_F$, and long enough junctions to ensure monotonic
decay of $I_c(L)$, $\xi_N^\prime$ appears to be shorter.

We use a Kubo linear response formalism to determine the normal
state resistance $R_N=1/G_N$. Kubo theory is formulated in terms
of the non-local conductivity tensor $\mbox{\b{$\sigma$}} ({\bf
r},{\bf r}^\prime;\nu)$
\begin{equation}\label{eq:locohm}
  {\bf j}(\bf{r},\nu) = \int \! d {\bf r}^\prime \, \mbox{\b{$\sigma$}}
  ({\bf r},{\bf r}^\prime;\nu) \cdot {\bf E}({\bf r}^\prime,\nu),
\end{equation}
which relates the current density ${\bf j}(\bf{r})$ to the
electric field ${\bf E}({\bf r})$ through a non-local Ohm's
law (at finite frequency $\nu$ these are the respective Fourier
components). Its physical meaning is obvious---it gives the
current response at ${\bf r}$ due to an electric field at ${\bf
r}'$. Although an external electric field induces charges (and
corresponding potentials) to linear order, the linear transport
properties, like $\mbox{\b{$\sigma$}} ({\bf r},{\bf r}^\prime)$,
are found as the response to an external field only. This is
because the current response to this inhomogeneous field
(external + induced) is already beyond linear
response.~\cite{nikolic_cpc} Thus, only equilibrium screening has
to be included in the Hamiltonian used to compute the Green
function entering $\mbox{\b{$\sigma$}} ({\bf r},{\bf r}')$
below.~\cite{stone} This makes it possible to use the potential
generated by the charge distributions $\delta n(z)$ (discussed in
Sec.~\ref{sec:lattice}), which is computed from the imaginary
axis calculations, as an on-site fixed potential in the
equilibrium Hamiltonian~(\ref{eq:tbh}). In this way
the potential of the SDLs enters the resistance calculation through the Green
functions in Eq.~(\ref{eq:kubonlct}) computed by real-axis
analytic continuation. The DC conductance of the sample of volume
$\Omega$ is expressed through $\mbox{\b{$\sigma$}} ({\bf r},{\bf
r}^\prime)$ as
\begin{eqnarray}\label{eq:volumecond}
  G_N & = & \frac{1}{V^2} \int\limits_{\Omega} d{\bf r} \, {\bf E}({\bf r})
  \cdot {\bf j}({\bf r})  \nonumber \\
  & = & \frac{1}{V^2} \int\limits_{\Omega}  d{\bf r} \, d{\bf r}' \, {\bf E}({\bf r})
  \cdot \mbox{\b{$\sigma$}} ({\bf r},{\bf r}') \cdot {\bf E}({\bf r}^\prime),
\end{eqnarray}
where ${\bf E}({\bf r})$ is the local field inside the sample and
$V$ is the externally applied voltage. Because of current
conservation requirements on the form of~\cite{kane}
$\mbox{\b{$\sigma$}} ({\bf r},{\bf r}^\prime)$, it is possible to
use arbitrary electric field factors in Eq.~(\ref{eq:volumecond})
[including a homogeneous field $E=V/L$].

Since our system is effectively one-dimensional (in real space)
we need to calculate the longitudinal component in the
$z$-direction (perpendicular to the uniform planes) of
$\mbox{\b{$\sigma$}} (z_i,z_i^\prime)$. In a lattice model like
ours, the relevant component of this tensor,
$\sigma_{zz}(\alpha,\beta)$, is given by (neglecting vertex
corrections)
\begin{eqnarray} \label{eq:kubonlct}
\sigma_{zz}(\alpha,\beta) & = & \frac{-1}{k_{B}T}
\frac{(eat)^2}{\hbar} \int\limits^{\infty}_{-\infty}
\rho^{2D}(\varepsilon_{xy}) d\varepsilon_{xy}
\int\limits^{\infty}_{-\infty} \frac{d\omega}{2\pi} \nonumber
\nonumber \\ & & \mbox{} \times [ \mbox{\large Im} \,
G_{\alpha,\beta+1}(\omega,\varepsilon_{xy})\, \mbox{\large Im}\,
G_{\beta,\alpha+1}(\omega,\varepsilon_{xy}) -
\nonumber \\
&& \mbox{\large Im}\,  G_{\alpha,\beta}(\omega,\varepsilon_{xy})
\, \mbox{\large Im}\,
G_{\beta+1,\alpha+1}(\omega,\varepsilon_{xy}) ]
 \nonumber \\
&& \mbox{} \times [\cosh^{2} (\omega/2k_{B}T)]^{-1},
\end{eqnarray}
where $f(\omega)$ is the Fermi-Dirac distribution function. We
first find the self-consistent solutions for the system in the normal
state, with no current flowing, by setting the order parameter to
zero on all planes. These solutions are then employed to
calculate the Kubo tensor~(\ref{eq:kubonlct}). The self-energy of
the planes outside the interlayer contains only a constant real
part, as the calculation is carried out within the HFA. Given the
set of local self-energies, the Green functions which couple any
two planes are readily found, for any momentum parallel to the
planes.

\begin{figure}
\centerline{\psfig{file=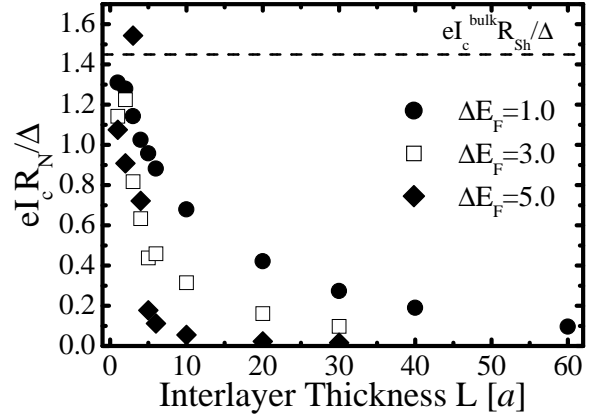,height=3.0in,angle=-90} }
\vspace{0.2in} \caption{Product of the critical current $I_c$ and
the normal state resistance $R_N$ as a function of the $SINIS$
junction thickness, for different SDLs determined by $\Delta
E_F=E_F^N - E_F^S$. Both the $S$ and $N$ are at half-filling in
the bulk. The $I_cR_N$ is always below the product of the bulk
critical current and Sharvin point contact resistance $I_c^{\rm
bulk} R_{\rm Sh}=1.45 \Delta/e$ (dashed line), except in the case
$L=3a (\approx l_D)$ with $\Delta E_F=5.0$.} \label{fig:IcRN}
\end{figure}

The conductance $G_N$ (per unit area $a^2$) of the lattice system
is obtained from the discretized version of~(\ref{eq:volumecond})
\begin{equation}
\frac{G_N}{a^2}=(R_{N}a^2)^{-1} = \sum_{\alpha,\beta}
\sigma_{zz}(\alpha,\beta),
\end{equation}
as the sum of the components of the non-local Kubo conductivity
tensor. Thus, although one can find the inhomogeneous
field~\cite{miller} $E_{\beta,\beta+1}$ (across all links
connecting planes $\beta$ and $\beta+1$) by inverting the
discretized version of Eq.~(\ref{eq:locohm}),
$I_{\alpha,\alpha+1}=a \sum_\beta \sigma_{zz}(\alpha,\beta)
E_{\beta,\beta+1}$ for $I_{\alpha,\alpha+1}$ a constant
throughout the system, the final expression for the conductance
does not contain this field. The normal state resistances
calculated in this framework are plotted in Fig.~\ref{fig:RN}. In
thin junctions and for large enough $\Delta E_F$, a charge
depletion layer arises inside the $N$ which leads to a
non-monotonic behavior of $R_N$ (e.g., $R_N$ increases sharply
for $L=2a$ and $\Delta E_F=3.0$, or $L=3a$ and $\Delta E_F=5.0$).
On the other hand, for small enough $\Delta E_F \lesssim 1.0$ the
conductance is only slightly changed from the Sharvin point contact
conductance~\cite{sharvin} of a ballistic $SNS$ junction per unit
area $a^2$, $R_Na^2=[(2e^2/h)(k_F^2 /4\pi)]^{-1} \approx 1.58
ha^2/2e^2$. Therefore, comparison of Fig.~\ref{fig:Ic_vs_L} and
Fig.~\ref{fig:RN} shows that the SDL depresses the current
substantially, while only weakly increasing the resistance. This
reduces the $I_c R_N$ product, plotted in Fig.~\ref{fig:IcRN},
thus showing that charge accumulation layers are detrimental to
junction performance in electronics circuits. This is further
confirmed by the fact that $I_cR_N$ in most of these junction is
below the product of the bulk critical current and the Sharvin
point contact resistance $I_c^{\rm bulk} R_{\rm Sh}=1.45
\Delta/e$, which is the upper limit of the characteristic voltage
in a clean $SNS$ weak link (the $SNS$ junction made of the same
$S$ leads as studied here, but with a dirty $N$ interlayer,
exhibits $I_c R_N
> I_c^{\rm bulk} R_{\rm Sh}$, for some range of
parameters~\cite{freericks}). Therefore, the SDL induced
scattering on a $SN$ boundary is one of the mechanisms which can
account for the low $I_cR_N$ products observed in
experiments~\cite{klapwijk} on nominally ballistic short $SNS$
junctions (where $R_N$, being determined by the thin charge layer
only, does not scale with $L$ just like what happens in ballistic
conductors). One way to test this conjecture is to use electron holography to map
out 
the charge  profile near the SN interface of
these ballistic junctions.

\section{Conclusions}
\label{sec:conclusion} We have studied the influence of a charge
imbalance, that arises at the boundary between a short coherence
length superconductor and a normal metal (due to Fermi energy
mismatch) on the equilibrium properties of a $SINIS$ Josephson
junction (where the $S$ and $N$ layers are of the same width). The
screening length is large enough to generate a spatially extended
charge redistribution  that allows us to examine the interplay
between the charge layer formation and superconductivity
(characterized by the coherence length comparable to the screening
length) near the $SN$ boundary and in the $N$ interlayer. This
resembles the charge redistribution on the grain boundaries of a
high-$T_c$ superconductor. At half-filling in both the $S$ and
$N$, the charge distribution and its potential are symmetric
(screened dipole layer), and can be rescaled to a single one
determined by some reference Fermi level mismatch $\Delta E_F$.
When charge concentration in the $N$ is a hundred times smaller than in 
the  $S$, we find a proximity effect induced change in the charge 
redistribution generated by a small Fermi level mismatch upon moving 
from $T > T_c$ (where $T$ is of the order of $\Delta E_F$) 
to $T<T_c$.

The step-function-like order parameter (which is used in
non-self-consistent approaches) is recovered only in the case of
a low charge density in the $N$ (compared to the filling in the $S$)
and a small mismatch $|E_F^N-E_F^S| \lesssim 1$. The $SINIS$ junction
exhibits unusual properties when its thickness is comparable to
the screening length. While the charge layer leads to a depression
of the order parameter near the $SN$ boundary, and thereby the
junction critical current, it influences the normal state resistance
in a much weaker fashion. Therefore, the $I_cR_N$ product, relevant for
digital electronics application, is reduced. This points out that
such space-charge layers should be avoided to optimize junction
performance and increase the critical current in high-$T_c$
superconductors.~\cite{mannhart99}

\section*{Acknowledgements}
We are grateful to the Office of Naval Research for financial
support from the grant number N00014-99-1-0328. Real-axis
analytic continuation calculations were partially supported by
HPC time from the Arctic Region Supercomputer Center. We have
benefited from the useful discussions with  A. Brinkman, J.
Ketterson, T. Klapwijk, K. K. Likharev, J. Mannhart, I.
Nevirkovets, N. Newman, I. V. Roshchin, J. Rowell, S. Tolpygo,
and T. van Duzer.



\begin{references}

\bibitem{josephson} B. D. Josephson, Phys. Lett. {\bf 1}, 251 (1962).

\bibitem{vanduzer} T. van Duzer and C. W. Turner, {\em Priciples of Superconducting
Devices and Circuits} (2nd ed., Prentice Hall, Upper Saddle River, 1999).

\bibitem{kleinssaser} A. Kastalsky, A.W. Kleinsasser, L.H. Greene, R. Bhat,
F.P. Milliken, and J.P. Harvison, Phys. Rev. Lett. {\bf 67}, 3026 (1991);
A. W. Kleinssaser and A. Kastalsky, Phys. Rev. B {\bf 47}, 8361 (1993).

\bibitem{brinkman} A. Brinkman and A. A. Golubov,
Phys. Rev. B {\bf 61}, 11 297 (2000).

\bibitem{volkov} A. F. Volkov, Phys. Rev. Lett. {\bf 74}, 4730
(1995).

\bibitem{carlo_rmt} C. W. J. Beenakker, Rev. of Mod. Phys. {\bf 69}, 731 (1997).

\bibitem{superlattices} Special issue of {\em Superlattices and
Microstructures}, {\bf 25} No. 5/6 (1999).

\bibitem{maezawa} M. Maezawa and A. Shoji, Appl. Phys. Lett. {\bf
70}, 3603 (1997).

\bibitem{sinis_review} For a review see: M. Yu. Kupriyanov, A. Brinkman, A. A.
Golubov, M. Siegel, H. Rogalla, Physica C {\bf 326-327}, 16
(1999).

\bibitem{rusi_sct} A. Brinkman, A. A. Golubov, H. Rogalla, and M.
Kupriyanov, Supercond. Sci. Technol. {\bf 12}, 893 (1999); D.
Balashov, F.-Im. Buchholz, H. Schulze, M. I. Khabipov, R. Dolata,
M. Yu. Kupriyanov, and J. Niemeyer, Supercond. Sci. Technol. {\bf
12}, 244 (2000).


\bibitem{andreev} A. F. Andreev, Zh. Eksp. Teor. Fiz. {\bf 46}, 1823 (1964) [Sov.
Phys. JETP {\bf 18}, 1228 (1964)].

\bibitem{likharev_review} K.~K.~Likharev, Rev. Mod. Phys. {\bf
51}, 101 (1979).

\bibitem{squid} M. B. Ketchen, IEEE Trans. Magn. {\bf 27}, 2916 (1991).

\bibitem{rsfq} K. K. Likharev and V. K. Semenov, IEEE Trans. Appl. Supercond.
{\bf 1}, 1 (1991).

\bibitem{gurvitch} M. Gurvitch, M. A. Washington, and H. A. Huggins, Appl.
Phys. Lett. {\bf 42}, 472 (1983).

\bibitem{klein_ieee} A. W. Kleinsasser, A. C. Callegari, B. D. Hunt, C. Rogers,
R. Tiberio, and R. A. Buhrman, IEEE Trans. Magn. {\bf 17}, 307
(1981).

\bibitem{zehnder} A. Zehnder, Ph. Lerch, S. P. Zhao, Th. Nussbaumer, E. C. Kirk, and
H. R. Ott, Phys. Rev. B {\bf 59}, 8875 (1999).

\bibitem{larkin} L. G. Aslamasov, A. I. Larkin, and Yu. N.
Ovchinnikov, Sov. Phys. JETP {\bf 28}, 171 (1969).

\bibitem{schon} W. Belzig, F.K. Wilhelm, C. Bruder, G. Sch\" on, and A.D.
Zaikin, Superlattices and Microstructures {\bf 25}, 1251 (1999).

\bibitem{lukichev} M. Yu. Kupriyanov and V. F. Lukichev, Sov.
Phys. JETP {\bf 67}, 1163 (1988).

\bibitem{blonder} G. E. Blonder, M. Tinkham, and T. M. Klapwijk, Phys. Rev. B {\bf 25},
4515 (1982).

\bibitem{furusaki} A. Furusaki, H. Takayanagi, and M. Tsukada, Phys. Rev. B {\bf 45},
10 563 (1992).

\bibitem{chrestin} A. Chrestin, T. Matsuyama, and U. Merkt, Phys.
Rev. B {\bf 59}, 498 (1994).

\bibitem{wendin} G. Johansson, E. N. Bratus', V. S. Shumeiko, and
G. Wendin, Phys. Rev. B {\bf 60}, 1382 (1999).

\bibitem{ivan} I. P. Nevirkovets and S. E. Shafranjuk, Phys. Rev.
B {\bf 59}, 1311 (1999).

\bibitem{gijs} M.A.M. Gijs and G.E.W. Bauer, Adv. Phys. {\bf
46}, 286 (1997), and references therein.

\bibitem{dugaev} V. K. Dugaev, V. I. Litvinov, and P. P. Petrov,
Phys. Rev. B {\bf 52}, 5306 (1995).

\bibitem{zaitsev} A. V. Zaitsev, Zh. Eksp. Teor. Fiz. {\bf 86}, 1742 [Sov. Phys. JETP {\bf 59},
1015 (1985)].

\bibitem{delin} K. A. Delin and A. W. Kleinsasser, Supercond. Sci. Technol.
{\bf 9}, 227 (1996).

\bibitem{mannhart98} J. Mannhart and H. Hilgenkamp, Appl. Phys. Lett. {\bf 73}, 265 (1998).

\bibitem{gurevich98} A. Gurevich and E. A. Pashitskii, Phys. Rev. B {\bf 57},
13 878 (1998).

\bibitem{igor} I. V. Roshchin, A. C. Abeyta, L. H. Greene, T. A. Tanzer, J. F. Dorsten,
P. W. Bohn, S.-W. Han, P. F. Miceli, and J. F. Klem, unpublished; I. V. Roshchin, Ph.D. thesis, 
{\em Electronic and optical properties of thin-film superconductors and superconductor-semiconductor
 interfaces}, Department of Physics, University of Illinois at Urbana-Champaign, Urbana (2000);  
L. H. Greene, J. F. Dorsten, I. V. Roschchin, A. C. Abeyta, T. A.
Tanzer, G. Kuchler, W. L. Feldmann, P. W. Bohn, Czech. J. Phys.
{\bf 46}, 3115 (1996).

\bibitem{klapwijk} J.P. Heida, B.J. van Wees, T.M. Klapwijk,
and G. Borghs, Phys. Rev. B {\bf 60}, 13 135 (1999), and
references therein.

\bibitem{levy} A. Levy-Yeyati, A. Mart$\acute{\i}$n-Rodero, and
F. J. Garc$\acute{\i}$a-Vidal, Phys. Rev. B {\bf 51}, 3743 (1995);
J. C. Cuevas, A. Mart$\acute{\i}$n-Rodero, and A. Levy Yeyati,
Phys. Rev. B {\bf 54}, 7366 (1996).


\bibitem{sols} F.~Sols and J.~Ferrer, Phys. Rev. B {\bf 49},
15913 (1994).


\bibitem{bagwell} R. A. Reidel, L.-F. Chang, and F. Bagwell, Phys. Rev. B {\bf 54}, 16 082 (1996).

\bibitem{annett} A. M. Martin and J. F. Annett, in Ref.~\onlinecite{superlattices}.


\bibitem{miller} P. Miller and J.~K.~Freericks, J.
Phys.: Condens. Matter. {\bf 13}, 3187 (2001).

\bibitem{dege} P. G. de Gennes, {\it Superconductivity of Metals and Alloys}
(Addison-Wesley, 1966).


\bibitem{pott} M. Potthoff and W. Nolting, Phys. Rev. B {\bf 59}, 2549 (1999).

\bibitem{freericks} J. K. Freericks, B. K. Nikoli\' c, and P.
Miller, cond-mat/0103067.

\bibitem{bardeen} J. Bardeen, Rev. Mod. Phys. {\bf 34}, 667 (1962).


\bibitem{deutscher} G. Deutscher and P.G. De Gennes, in
{\em Superconductivity}, ed. by R.D. Parks, (Marcel Dekker, New
York, 1969), Vol. II, p. 1005.

\bibitem{falk} D. S. Falk, Phys. Rev. {\bf 132}, 1576 (1963).

\bibitem{lesovik} G. B. Lesovik, T. Martin, and G. Blatter,
cond-mat/0009193.


\bibitem{kupriyanov92} M. Yu. Kupriyanov, Pis'ma Zh. Eksp. Teor.
Fiz. {\bf 56}, 414 (1992) [JETP Lett. {\bf 56}, 399 (1992)].

\bibitem{jackson} A. W. Kleinsasser and T. N. Jackson, Phys. Rev. B {\bf 42},
R8716 (1990).

\bibitem{kane} C. L. Kane, R. A. Serota, and P. A. Lee, Phys. Rev. B {\bf 37},
6701 (1988).

\bibitem{nikolic_cpc} B. K. Nikoli\' c and P. B. Allen, Phys. Rev. B {\bf 60}, 3963 (1999).

\bibitem{stone} A. D. Stone, in \emph{Mesoscopic Quantum Physics}, edited by E. Akkermans, J.-L. Pichard, and J. Zinn-Justin, Les Houches, Session LXI, 1994 (North-Holland, Amsterdam, 1995)..

\bibitem{sharvin} Yu. V. Sharvin,  Zh. Eksp. Teor. Phys. {\bf 48}, 984 (1965)
[Sov. Phys. JETP {\bf 21}, 655 (1965)].

\bibitem{mannhart99} A. Scheml, B. Goetz, R. R. Schulz, C. W. Schneider, H. Bielefeldt, H. Hilgenkamp, and J. Mannhart, Europhys. Lett. {\bf 47}, 110 (1999); G. Hammerl,
A. Schmehl, R. R. Schulz, B. Goetz, H. Bielefeldt, C. W. Schneider,
H. Hilgenkamp, and J. Mannhart, Nature {\bf 407}, 162 (2000).

\end{references}
\end{document}